\documentclass[12pt]{article}
\pdfoutput=1

\usepackage{putex}
\usepackage{graphicx}
\usepackage{caption}
\usepackage{amsmath}
\usepackage{array}
\usepackage{subcaption}
\usepackage{epstopdf}
\usepackage{enumerate}
\usepackage{cite}
\usepackage{youngtab}
\usepackage{tensor}
\usepackage{slashed}
\usepackage[aligntableaux=center]{ytableau}
\usepackage[utf8]{inputenc}
\usepackage{rotating}
\usepackage{bigfoot}
\usepackage[
      colorlinks=true,
      linkcolor=blue,
      urlcolor=blue,
      filecolor=black,
      citecolor=red,
      ]{hyperref}

\newcommand {\be} {\begin {equation}}
\newcommand {\ee} {\end {equation}}

\newcommand {\no} {\nonumber}
\newcommand {\bes} {\begin {equation*}}
\newcommand {\ees} {\end {equation*}}

\newcommand{\es}[2] {\begin{equation} \label{#1} \begin{split} #2 \end{split} \end{equation}}
\def\g{\gamma}

\newcommand{\cA}{{\mathcal A}}
\newcommand{\cB}{{\mathcal B}}

\newcommand{\cD}{{\mathcal D}}

\newcommand{\cF}{{\mathcal F}}
\newcommand{\cG}{{\mathcal G}}
\newcommand{\cH}{{\mathcal H}}

\newcommand{\cL}{{\mathcal L}}
\newcommand{\cN}{{\mathcal N}}
\newcommand{\cO}{{\mathcal O}}
\newcommand{\cP}{{\mathcal P}}

\newcommand{\cT}{{\mathcal T}}

\newcommand{\cM}{{\mathcal M}}

\newcommand{\beq}{\begin{equation}}
\newcommand{\eeq}{\end{equation}}

\def\ie{\begin{equation}\begin{aligned}}
\def\fe{\end{aligned}\end{equation}}

\numberwithin{equation}{section}

\newcommand\Tstrut{\rule{0pt}{2.3ex}}       
\newcommand\Bstrut{\rule[-1.3ex]{0pt}{0pt}} 
\newcommand\TBstrut{\Tstrut\Bstrut}         

\def\<{\langle}
\def\>{\rangle}

\begin{document}

\preprint{}

\institution{oxford}{Mathematical Institute, University of Oxford,
Woodstock Road, Oxford, OX2 6GG, UK}
\institution{Exile}{Department of Particle Physics and Astrophysics, Weizmann Institute of Science, Rehovot, Israel}

\title{6d (2,0) and M-theory at 1-loop}

\authors{Luis F. Alday\worksat{\oxford}, Shai M. Chester\worksat{\Exile}, and Himanshu Raj\worksat{\Exile}}

\abstract{
We study the stress tensor multiplet four-point function in the 6d maximally supersymmetric $(2,0)$ $A_{N-1}$ and $D_N$ theories, which have no Lagrangian description, but in the large $N$ limit are holographically dual to weakly coupled M-theory on $AdS_7\times S^4$ and $AdS_7\times S^4/\mathbb{Z}_2$, respectively. We use the analytic bootstrap to compute the 1-loop correction to this holographic correlator coming from Witten diagrams with supergravity $R$ and the first higher derivative correction $R^4$ vertices, which is the first 1-loop correction computed for a non-Lagrangian theory. We then take the flat space limit and find precise agreement with the corresponding terms in the 11d M-theory S-matrix, some of which we compute for the first time using two-particle unitarity cuts. 
}
\date{}

\maketitle

\tableofcontents

\section{Introduction}
\label{intro}

The maximally supersymmetric $(2,0)$ conformal field theories (CFTs) in 6d are the most fundamental CFTs. No supersymmetric CFT can be defined in any higher dimension \cite{Nahm:1977tg}, and in fact no non-supersymmetric CFTs have been found in 6d or higher. Many lower dimensional CFTs can then be derived from the 6d $(2,0)$ theories, and the 6d origin can be used to derive deep non-perturbative aspects of these lower dimensional CFTs \cite{Gaiotto:2009we,Gaiotto:2009hg,Dimofte:2011ju,Bah:2012dg,Gadde:2013sca}. The $(2,0)$ theories are hard to study, however, because they have no Lagrangian, and were originally only defined by considering a decoupling limit of Type IIB string theory on ALE spaces \cite{Witten:1995zh}, which predicts the existence of a list of interacting $(2,0)$ theories labelled by Lie algebras $A_{N-1}$, $D_N$, $E_6$, $E_7$, $E_8$, in addition to a free $(2,0)$ theory.\footnote{For small $N$, this list is redundant because $D_1=SO(2)$ is the free theory, $D_3=A_3$, and $D_2=A_1\times A_1$, so this last theory is really a product theory with two stress tensors.} Since these theories have no Lagrangian, the powerful non-perturbative methods of integrability and supersymmetric localization cannot be used to study correlators.\footnote{The known $(2,0)$ theories can be conjecturally related to 5d SYM by compactifying on a circle \cite{Seiberg:1997ax,Douglas:2010iu,Lambert:2010iw,Lambert:2012qy,Seiberg:1996bd,Mezei:2019aa}, then localization can be applied to the 5d theory and used to study protected quantities in the 6d theory that are invariant to the radius of the circle. However, it is difficult to constrain the 6d correlator using this method.}  The only perturbative way of studying 6d correlators is using the holographic duality between the $A_{N-1}$ and $D_N$ theories at large $N$ and weakly coupled M-theory on $AdS_7\times S^4$ \cite{Maldacena:1997re} and $AdS_7\times S^4/\mathbb{Z}_2$ \cite{Witten:1998xy,Aharony:1998rm}, respectively. In particular, this duality maps four-point functions of single trace half-BPS operators in the CFT to scattering of gravitons and higher KK modes in AdS, which in the flat space limit becomes the 11d M-theory S-matrix $\cA$. The weak coupling expansion for $\cA$ consists of the small Planck length $\ell_{11}$, i.e. small momentum, expansion:
\es{A}{
\mathcal{A}(s,t)=\ell_{11}^{9}\mathcal{A}_{R}+\ell^{15}_{11}\mathcal{A}_{R^4}+\ell^{18}_{11}\mathcal{A}_{R|R}+\ell^{21}_{11}\mathcal{A}_{D^6R^4}+\ell^{23}_{11}\mathcal{A}_{D^8R^4}+\ell^{24}\mathcal{A}_{R|R^4}+\dots\,,
}
where $s,t,u$ are 11d Mandelstam variables. The lowest few tree level\footnote{Since $\ell_{11}$ is the only expansion parameter, there is in general no difference between tree and loop level. At low orders, however, one can distinguish between tree and loop terms by the different power of $\ell_{11}$ that multiply them.} terms $\cA_R$, $\cA_{R^4}$, and $\cA_{D^6R^4}$ are protected, and so can be computed from Type IIA string theory by compactifying on a circle \cite{Green:1997as, Russo:1997mk, Green:2005ba}\footnote{$\cA_{D^4R^4}$ can also be computed in this way, but it vanishes and so we did not write it.} to get
  \es{SGtoR4}{
  \frac{ \mathcal{A}_{R^4}}{\mathcal{A}_{R}}=\frac{stu}{3\cdot 2^7}\,,\qquad  \frac{ \mathcal{A}_{D^6R^4}}{\mathcal{A}_{R}}=\frac{(stu)^2}{15\cdot  2^{15}}\,.
   }
   The 1-loop supergravity term $\cA_{R|R}$ has also been computed in \cite{Russo:1997mk,Green:1997as}. The small $\ell_{11}$ expansion in 11d maps to the large $N$ expansion in the CFT according to the dictionary
   \cite{Maldacena:1997re,Intriligator:2000eq}:
\es{cPlanck}{
&(L_\text{AdS}/\ell_{11})^9\approx \mathfrak{o}16c\,,\\
}
where the orbifold factor $\mathfrak{o}$ is one/two for the $A_{N-1}$/$D_N$ theories, and $c$ is the central charge appearing in the (2,0) conformal anomaly, which for the $A_{N-1}$ and $D_N$ theories is\footnote{More generally, a $(2,0)$ theory labeled by Lie algebra $\mathfrak{g}$ has central charge $c(\mathfrak{g})=4d_\mathfrak{g}h^\vee_\mathfrak{g}+r_\mathfrak{g}$, where $d_\mathfrak{g}$, $h^\vee_\mathfrak{g}$, and $r_\mathfrak{g}$ are the dimension, dual Coxeter number, and rank of $\mathfrak{g}$, respectively \cite{Beem:2015aoa}.} \cite{Beem:2015aoa,Intriligator:2000eq,Cordova:2015vwa,Ohmori:2014kda,Cordova:2016cmu}
\es{cAD}{
c(A_{N-1})=4N^3-3N-1\,,\qquad c(D_{N})=16N^3-24N^2+9N\,.
}
One can then try to study the large $c\sim N^3$ expansion in the CFT from the small $\ell_{11}$ expansion in M-theory. This is difficult, however, since nothing is known about the M-theory S-matrix aside from the terms described above. This is unlike the paradigmatic holographic duality between $\cN=4$ SYM and Type IIB on $AdS_5\times S^5$, where the string theory S-matrix can be computed using the worldsheet to all orders in the string length to the lowest few orders in the string coupling \cite{Gomez:2010ad,Gomez:2013sla,DHoker:2005jhf}, and so can in principle be used to constrain the CFT to many orders.

Progress can be made by exploiting constraints from superconformal symmetry itself. The recent reformulation \cite{Rattazzi:2008pe} of the conformal bootstrap \cite{Polyakov:1974gs,Ferrara:1973yt,Mack:1975jr} uses crossing symmetry and semidefinite programming to place numerical but rigorous non-perturbative bounds on the scaling dimensions and operator-product-expansion (OPE) coefficients that appear in CFT correlators. The numerical bootstrap was applied to the stress tensor multiplet correlator of 6d $(2,0)$ CFTs in \cite{Beem:2015aoa}, and used to compute non-perturbative bounds on CFT data as a function of $c$. In particular, \cite{Beem:2015aoa} showed evidence that the leading $c^{-1}$ correction to various CFT data, as computed from M-theory in the supergravity limit \cite{Heslop:2004du,Arutyunov:2002ff}, saturated the bounds at large $c$, and that the lowest known interacting theory $A_1$ saturated the lower bound on $c$. However, the numerical bootstrap is much less numerically stable in 6d than in lower dimensions, so it was difficult to make a precise comparison to analytical results. Also, the bootstrap bounds in principle apply to any theory with a given value of $c$, but one cannot focus on the physical 6d CFTs, such as the $A_{N-1}$ and $D_N$ theories, unless one conjectures that these theories saturate the bounds. The $D_N$ and $A_{N-1}$ theories are identical at order $c^{-1}$, since they are both described by the same supergravity, so even with infinite precision it would be impossible to see which theory was saturating the bound by comparing to this analytic result.

Instead of exploiting crossing symmetry numerically for general CFTs at finite $c$, one can apply it to the large $c$ expansion of specific holographic theories. This analytic bootstrap uses the analytic structure of Witten diagrams in Mellin space, the flat space limit, and crossing symmetry to constrain holographic correlators in a large $c$ expansion. For 6d $(2,0)$ theories, these constraints were used to compute the stress tensor correlator at leading order in $c^{-1}$ in position space \cite{Arutyunov:2002ff,Heslop:2004du}, and then extended to Mellin space in \cite{Rastelli:2017ymc} for the correlators $\langle pppp\rangle$ for $p=2,3,4$, where $p$ denotes the bottom component of the $p$th lowest single trace half-BPS multiplet (e.g. $p=2$ corresponds to the stress tensor multiplet), which are dual to the corresponding $p$th lowest scalar KK mode in the dimensional reduction of M-theory on AdS. In \cite{Chester:2018dga}, the analytic bootstrap was then combined with the constraints from the protected 2d chiral algebra subsector \cite{Beem:2014kka} for the $A_{N-1}$ theory to compute the first higher derivative tree level correction to $\langle 3333\rangle$ coming from the $R^4$ term in the AdS effective action, which contributes as $c^{-\frac53}$.

In this paper we will compute the first 1-loop corrections at large $c$ to $\langle2222\rangle$ coming from Witten diagrams with supergravity $R$ and higher derivative $R^4$ vertices for both the $A_{N-1}$ and $D_N$ theories. In particular, we will compute the $R|R$ term at $c^{-2}$, the $R|R^4$ term at $c^{-\frac83}$, and the $R^4|R^4$ term at $c^{-\frac{10}{3}}$. As shown in \cite{Aharony:2016dwx}, the 1-loop correction to holographic correlators can in general be computed  by ``squaring'' the tree level anomalous dimensions of all double-trace operators in the correlator. More precisely, from this tree level data one can compute the double-discontinuity using crossing symmetry, from which the full correlator can be reconstructed as in \cite{Alday:2016njk,Caron-Huot:2017vep}. For known holographic theories, these double trace operators are degenerate in the generalized free field theory (GFFT) that describe the strict $c\to\infty$ limit, and so their tree level anomalous dimensions at orders $c^{-1}$ for $R$ and $c^{-\frac53}$ for $R^4$ must be unmixed to obtain the 1-loop corrections we consider. This unmixing requires the average of GFFT OPE coefficients obtained from $\langle pppp\rangle$, as well as the average of $R$ and $R^4$ anomalous dimensions obtained from $\langle 22pp\rangle$. For $\mathcal{N}=4$ SYM, the GFFT and $R$ data was computed using conformal symmetry \cite{Dolan:2004mu,Dolan:2001tt,Alday:2014qfa,Rastelli:2017udc}, and then used to obtain the 1-loop correction in \cite{Alday:2017xua,Aprile:2017xsp,Aprile:2017bgs,Alday:2018kkw}. The $R^4$ data was computed from comparing to the known Type IIB $R^4$ term in the flat space limit \cite{Penedones:2010ue,Goncalves:2014ffa,Alday:2018pdi}\footnote{They were later independently computed from supersymmetric localization in \cite{Binder:2019jwn}, which was also used to compute higher order terms in \cite{Chester:2019pvm,Chester:2020dja,Chester:2019jas}.}, and then used to obtain the 1-loop correction in \cite{Alday:2018pdi}.\footnote{See \cite{Alday:2019nin,Drummond:2019odu,Drummond:2019hel,Aprile:2019rep,Aprile:2018efk,Aprile:2017qoy,Drummond:2020dwr,Caron-Huot:2018kta,Alday:2017vkk} for generalizations to other 1-loop correlators and \cite{Bissi:2020wtv} for progress at two loops in $\mathcal{N}=4$ SYM.} 

We carry out a similar program for both the $A_{N-1}$ and $D_N$ theories, where the only difference between each theory in this calculation is that the orbifold for $D_N$ projects out all odd $p$ multiplets \cite{Aharony:1998rm},\footnote{This statement is strictly only true in the large $N$ limit. At finite $N$, the $D_N$ theories have an extra Pfaffian operator with $p=N$\cite{Aharony:1998rm} which can be odd.} and the $R^4$ data comes with the orbifold factor $\mathfrak{o}$ in \eqref{cPlanck}. For the tree $\langle 22pp\rangle$ data, we could easily extract the average anomalous dimension from the $R$ correlator given in \cite{Zhou:2017zaw} as well as the $R^4$ correlator that we compute here using the flat space limit and \eqref{SGtoR4}. Far more challenging were the average GFFT OPE coefficients from $\langle pppp\rangle$, for which we needed to compute the full superconformal block expansion up to $p=11$ using the superconformal Ward identities in \cite{Dolan:2004mu}, and then guess the general $p$ formula. We then use this to compute $R|R$, $R|R^4$ and $R^4|R^4$ in both the $A_{N-1}$ and $D_N$ theories, extract the low-lying CFT data,\footnote{We did not do this for $R^4|R^4$, since at order $c^{-\frac{10}{3}}$ it is subleading to the unknown 2-loop term $R|R|R$, but this extraction could easily be done using our methods.} and compare to the M-theory S-matrix in the flat space limit. For $R|R$, we precisely match the 1-loop correction to 11d supergravity \cite{Russo:1997mk,Green:1997as}, as was done for $\mathcal{N}=4$ SYM and 10d supergravity in \cite{Alday:2019nin}. For $R|R^4$ and $R^4|R^4$, we similarly compute the correlators in both theories and extract CFT data. The corresponding 11d terms $\cA_{R|R^4}$ and $\cA_{R^4|R^4}$ were not computed previously, but we derive them from the known tree level terms using two-particle unitarity, as was originally done for 10d string theory in \cite{Green:2008uj}. We then find a precise match with the 6d holographic correlator at these orders in the flat space limit, which is the first presice check of AdS/CFT for 1-loop terms with higher derivative vertices.\footnote{Similar 1-loop term were considered for $\cN=4$ SYM in \cite{Alday:2018pdi} and compared to the corresponding 10d terms as computed from two-particle unitarity in \cite{Green:2008uj}, but the authors were only able to match the ratio $\cA_{R|R^4}/\cA_{R^4|R^4}$ using the unjustified assumption of $\alpha'=4$.}

 The rest of this paper is organized as follows. In Section~\ref{4points} we review the constraints of superconformal symmetry on $\langle qqpp\rangle$ for $q\leq p$, and compute the explicit superblock decomposition for $p\leq11$. In Section~\ref{largecc}, we extract the average GFFT OPE coefficients from $\langle pppp\rangle$ and the average tree level anomalous dimensions from $\langle 22pp\rangle$, and discuss the large $c$ expansion of $\langle2222\rangle$. In Section~\ref{1loop}, we use this data to compute the 1-loop corrections to $\langle2222\rangle$, compare to the 1-loop terms in the M-theory S-matrix in the flat space limit, and extract CFT data from the 1-loop correlators. Finally, in Section~\ref{conc} we end with a discussion of our results and of future directions.  Several technical details, including the two-particle unitarity cut formula for 11d, are given in various Appendices, and we include an attached \texttt{Mathematica} notebook with explicit results.

\section{Half-BPS four-point functions in 6d}
\label{4points}

We begin by discussing the constraints of the 6d $(2,0)$ superconformal algebra $\mathfrak{osp}(8^*|4)\supset\mathfrak{so}(2)\oplus \mathfrak{so}(6)\oplus\mathfrak{so}(5)_R$ on four-point functions of the dimension $2p$ scalar bottom component of half-BPS supermultiplets in an interacting theory. First we discuss the $s$-channel superconformal block expansion for $\langle qqpp\rangle$ correlators.  Then we focus on $\langle 22pp\rangle$, which has a simpler superblock expansion in terms of a reduced correlator that solves the superconformal Ward identity. The analysis of this section applies to any interacting 6d $(2,0)$ CFT.

\subsection{Block expansion of $\langle qqpp\rangle$}
\label{qqpp}

We consider half-BPS superconformal primaries $S_p$ in 6d $(2,0)$ SCFTs that are scalars with $\Delta=2p$ and transform in the $[p\,0]$ of $\mathfrak{so}(5)_R$, where $p=1,2,\dots$. The lowest case $S_1$ is a free scalar with $\Delta=2$ in the ${\bf5}$ of $\mathfrak{so}(5)_R$. We are interested in interacting theories, so we will restrict to $p>1$. The first such interacting operator is $S_2$, which is the bottom component of the conserved stress tensor multiplet and so must exist in all local 6d $(2,0)$ SCFTs. We can view these operators as the rank-$p$ symmetric traceless products of the ${\bf5}$, so we can denote them as traceless symmetric tensors $S_{I_1\dots I_p}( x)$ of $\mathfrak{so}(5)_R$, where $I_i=1,\dots5$. It is convenient to contract with an auxiliary polarization vector $Y^I$ that is constrained to be null, $Y_i\cdot Y_i=0$, so that
\es{S}{
S_p( x,Y)\equiv S_{I_1\dots I_p}Y^{I_1}\cdots Y^{I_p}\,.
}

We are interested in four-point functions $\langle qqpp\rangle$ of four $S_p(x,Y)$'s, where we take $q\leq p$. These correlators are fixed by conformal and $\mathfrak{so}(5)_R$ symmetry to take the form
\es{4point}{
\langle S_q( x_1,Y_1)S_q( x_2,Y_2)S_p( x_3,Y_3)S_p( x_4,Y_4) \rangle=\frac{(Y_1\cdot Y_2)^q(Y_3\cdot Y_4)^p}{| x_{12}|^{4q}| x_{34}|^{4p}}\mathcal{G}_{qp}(U,V;\sigma,\tau)\,,
}
where $U$ and $V$ are conformally-invariant cross ratios and $\sigma$ and $\tau$ are $\mathfrak{so}(5)_R$ invariants formed out of the polarizations:
 \es{uvsigmatauDefs}{
  U \equiv \frac{{x}_{12}^2 {x}_{34}^2}{{x}_{13}^2 {x}_{24}^2} \,, \qquad
   V \equiv \frac{{x}_{14}^2 {x}_{23}^2}{{x}_{13}^2 {x}_{24}^2}  \,, \qquad
   \sigma\equiv\frac{(Y_1\cdot Y_3)(Y_2\cdot Y_4)}{(Y_1\cdot Y_2)(Y_3\cdot Y_4)}\,,\qquad \tau\equiv\frac{(Y_1\cdot Y_4)(Y_2\cdot Y_3)}{(Y_1\cdot Y_2)(Y_3\cdot Y_4)} \,,
 }
 with $x_{ij}\equiv x_i-x_j$. Since \eqref{4point} is a degree $q$ polynomial in each $Y_i$ separately, the quantity $\mathcal{G}_{qp}(U,V;\sigma,\tau)$ is a degree $q$ polynomial in $\sigma$ and $\tau$. It is convenient to parametrize these polynomials in terms of eigenfunctions $Y_{mn}(\sigma, \tau)$ of the $\mathfrak{so}(5)_R$ quadratic Casimir for irreps $[ 2n\, 2(m-n)]$ that appear in the tensor product of $[q\,0]\otimes[q\,0]$, so that $m=0,1,\dots q$ and $n=0,\dots m$ .  The polynomials $Y_{mn}(\sigma, \tau)$ can be computed explicitly as shown in Appendix D of \cite{Nirschl:2004pa}; we give the explicit values for $q=2,\dots 11$ in the attached \texttt{Mathematica} file. We can then expand $\mathcal{G}_{qp}(U,V;\sigma,\tau)$ in terms of this basis as
 \es{Ybasis}{
\mathcal{G}_{qp}(U,V;\sigma,\tau)=\sum^{q}_{m=0}\sum_{n=0}^{m}Y_{mn}(\sigma, \tau) A_{mn}(U,V)\,.
}
By taking the OPEs $S_{q}\times S_{q}$ and $S_{p}\times S_{p}$ in \eqref{4point}, we can expand $A_{mn}(U,V)$ in the $s$-channel conformal blocks $G_{\Delta,\ell}(U,V)$ as
\es{blockExp}{
A_{mn}(U,V)=\sum_{\Delta,\ell}\lambda_{qq\cO_{\Delta,\ell,mn}}\lambda_{pp\cO_{\Delta,\ell,mn}}G_{\Delta,\ell}(U,V)\,,
}
where $\cO_{\Delta,\ell,nm}$ are conformal primaries with scaling dimension $\Delta$ and spin $\ell$ in irrep $[ 2n\, 2(m-n)]$ that appear in $S_{q}\times S_{q}$ with OPE coefficient $\lambda_{qq\cO_{\Delta,\ell,mn}}$. The 6d conformal blocks were originally computed in closed form in \cite{Dolan:2011dv,Dolan:2003hv}, which we show explicitly in our conventions in Appendix \ref{block6dApp}.

So far, we have imposed the bosonic subgroups of the $\mathfrak{osp}(8^*|4)$ algebra. The constraints from the fermionic generators are captured by the superconformal Ward identities \cite{Dolan:2004mu}:
\es{ward}{
\left[z\partial_z -  2\alpha \partial_\alpha\right] \mathcal{G}_{qp}(z,\bar{z};\alpha, \bar{\alpha})|_{\alpha = \frac1z} =
\left[\bar{z}\partial_{\bar{z}} -  2{\bar\alpha} \partial_{\bar{\alpha}}\right] \mathcal{G}_{qp}(z,\bar{z};\alpha, \bar{\alpha})|_{\bar{\alpha}=\frac{1}{\bar{z}}} &= 0\,,
}
where $z,\bar z$ and $\alpha,\bar\alpha$ are written in terms of $U,V$ and $\sigma,\tau$, respectively, as
\es{UVtozzbar}{
U=z\bar z\,,\quad V=(1-z)(1-\bar z)\,,\qquad\qquad \sigma=\alpha\bar\alpha\,,\quad \tau=(1-\alpha)(1-\bar\alpha)\,.
}
These constraints can be satisfied by expanding $ \mathcal{G}_{qp}$ in superconformal blocks as
\es{SBDecomp}{
      \mathcal{G}_{qp}(U,V;\sigma,\tau)=\sum_{\mathcal{M}\in S_{q}\times S_{q}}\lambda_{qq\mathcal{M}} \lambda_{pp\mathcal{M}} \mathfrak{G}_{\mathcal{M}}(U,V;\sigma,\tau)\,,
}
where $\mathfrak{G}_{\mathcal{M}}$ are superblocks for each supermultiplet $\mathcal{M}$ that appears in $S_{q}\times S_{q}$ (and $S_{p}\times S_{p}$)  with OPE coefficients $\lambda_{qq\mathcal{M}}$ (and $\lambda_{pp\mathcal{M}}$). The selection rules for the OPE of half-BPS multiplets have been worked out in \cite{Heslop:2004du,Ferrara:2001uj} and were summarized for general $q$ in \cite{Beem:2015aoa}. The supermultiplets that appear in $S_q\times S_q$ are
\es{opemultEq}{
&S_q\times S_q=\sum_{m=0}^q\sum_{n=0}^{q-m}\mathcal{D}[2(q-m-n),2n]\\
&+\sum_{m=1}^q\left[\sum_{n=0,2,\dots}^{q-m}\,\sum_{\ell=0,2,\dots}^\infty\mathcal{B}[2(q-m-n),2n]_\ell+\sum_{n=1,3,\dots}^{q-m}\,\sum_{\ell=1,3,\dots}^\infty\mathcal{B}[2(q-m-n),2n]_\ell\right]\\
&+\sum_{m=2}^q\left[\sum_{n=0,2,\dots}^{q-m}\sum_{\substack{\ell=0,2,\dots\\ \Delta}}^\infty\mathcal{L}[2(q-m-n),2n]_{\Delta,\ell}+\sum_{n=1,3,\dots}^{q-m}\sum_{\substack{\ell=1,3,\dots\\ \Delta}}^\infty\mathcal{L}[2(q-m-n),2n]_{\Delta,\ell}\right],\\
}
where the spins $\ell$ refer to rank-$\ell$ traceless symmetric irreps of the Lorentz group $\mathfrak{so}(6)$ with Dynkin labels $[\ell00]$, which are the only irreps that can appear, and for interacting SCFTs we should further remove the $\mathcal{B}[00]_\ell$ multiplet, which contains higher spin conserved currents that only appear in the free theory. The scaling dimensions of bottom components of the supermultiplets in \eqref{opemultEq} are
\es{dims}{
&\mathcal{D}[p_1p_2]:\qquad\, \Delta=2(p_1+p_2)\,,\\
&\mathcal{B}[p_1p_2]_\ell:\qquad\Delta=2(p_1+p_2)+4+\ell\,,\\
&\mathcal{L}[p_1p_2]_{\Delta,\ell}:\quad\, \Delta>2(p_1+p_2)+6+\ell\,.\\
}
The $\mathcal{L}$ multiplets that appear here are unprotected, while the rest are annihilated by some fraction of supercharges and so have fixed dimension. The $\mathcal{D}[p0]$ are the half-BPS multiplets whose bottom component we called $S_p$, and $q$ of these multiplets appear in $S_q\times S_q$. The lowest such multiplet is always the stress tensor multiplet $\mathcal{D}[20]$, whose OPE coefficient squared is fixed by the conformal Ward identity \cite{Osborn:1993cr} to be inversely proportional to the coefficient of the canonically normalized stress tensor two-point function:
\es{stress}
{ \langle T_{\mu\nu}(x)T_{\rho\sigma}(0) \rangle = c\frac{84}{\pi^6} {{\cal I_{\mu\nu\rho\sigma}}(x)\over | x|^{12}}\,,
}
where ${\cal I}_{\mu\nu\rho\sigma}(x)$ is a fixed tensor structure whose form can be found in \cite{Osborn:1993cr}, and $c=1$ for the free theory. In this normalization we get the precise relationship
\es{cTolam}{
\lambda^2_{pp\mathcal{D}[20]}=\frac{p^2}{{c}}\,.
}

We can now compare \eqref{SBDecomp} to \eqref{Ybasis} and \eqref{blockExp} to see that the superblocks are finite linear combinations of conformal blocks 
\es{GExpansion}{
\mathfrak{G}_{\mathcal{M}}=\sum^{q}_{m=0}\sum_{n=0}^{m}Y_{mn}(\sigma, \tau)  \sum_{\cO\in\mathcal{M}} \frac{\lambda_{qq\cO_{\Delta,\ell,mn}} \lambda_{pp\cO_{\Delta,\ell,mn}}   }{ \lambda_{qq\mathcal{M}} \lambda_{pp\mathcal{M}}  }G_{\Delta,\ell}(U,V)\,,
}
where $\cO_{\Delta,\ell,mn}$ are conformal primaries that appear in $\mathcal{M}$, which can be computed using the Racah-Speiser algorithm in \cite{Buican:2016hpb,Cordova:2016emh}. The allowed $\cO_{\Delta,\ell,mn}$ are further restricted to have $\mathfrak{so}(5)_R$ irreps that can appear in $[q\,0]\otimes[q\,0]$, and have even/odd $\ell$ for irreps appearing in the symmetric/antisymmetric product, where $[2a,2b]$ are symmetric (antisymmetric) when $b$ is even (odd). For $\langle qqpp\rangle$, all of the pairs of OPE coefficients $\lambda_{qq\cO_{\Delta,\ell,nm}}\lambda_{pp\cO_{\Delta,\ell,nm}}$ can be fixed in terms of the single pair of OPE coefficients $\lambda_{qq\mathcal{M}} \lambda_{pp\mathcal{M}} $ in \eqref{SBDecomp} using the Ward identities \eqref{ward}. In practice, following a very similar analysis done for 3d $\mathcal{N}=8$ in \cite{Chester:2014fya,Agmon:2017xes}, we expanded the conformal blocks for the conformal primaries that appear in a given multiplet $\mathcal{M}$ using the small $z,\bar z$ expansion in terms of Jack polynomials \cite{Dolan:2003hv}, as we review in Appendix \ref{block6dApp}, and then applied the Ward identities order by order to fix the coefficients $\frac{\lambda_{qq\cO_{\Delta,\ell,mn}} \lambda_{pp\cO_{\Delta,\ell,mn}}   }{ \lambda_{qq\mathcal{M}} \lambda_{pp\mathcal{M}}  }$. This is a tedious but conceptually straightforward exercise, and we give the solutions for all the superblocks that appear in the $s$-channel of $\langle qqpp\rangle$ for $q=2,\dots,11$ in the attached \texttt{Mathematica} file. 

\subsection{Reduced correlator for $\langle 22pp\rangle$}
\label{22pp}

For $\langle22pp\rangle$, there is a different way of satisfying the superconformal Ward identities as shown in \cite{Dolan:2004mu} that we will find particularly useful. We do this by writing $\mathcal{G}_{2p}$ as\footnote{In fact, this formal solution to the Ward identities exists for any $\langle qqpp\rangle$, but a superblock expansion of the reduced correlators was only worked out for the case $\langle 2222\rangle$ \cite{Beem:2015aoa}, which we will generalize here to $\langle22pp\rangle$.}
\es{Ward2}{
\mathcal{G}_{2p}(U,V;\sigma,\tau)=\mathcal{F}_{p}(U,V;\sigma,\tau)+\Upsilon\circ \mathcal{H}_{p}(U,V)\,,
}
where the reduced correlator $\mathcal{H}_{p}(U,V)$ is an arbitrary function of degree two and zero in $\sigma,\tau$, respectively, and $\Upsilon$ is the differential operator given in \cite{Dolan:2004mu,Rastelli:2017ymc} as
\es{Y}{
\Upsilon&=\sigma^2\mathcal{D}'UV+\tau^2\mathcal{D}'U+\mathcal{D}'V-\sigma\mathcal{D}'V(U+1-V)
-\tau\mathcal{D}'(U+V-1)-\sigma\tau\mathcal{D}'U(V+1-U)\,,\\
\mathcal{D}'&=D-\frac{2}{V}(D^+_0-D^+_1+2\partial_\sigma \sigma)\tau\partial_\tau+\frac{2}{UV}(-V D_1^++2(V\partial_\sigma \sigma+\partial_{\tau}\tau-1))(\partial_\sigma\sigma+\partial_\tau\tau)\,,\\
D&=\partial_z\partial_{\bar z}-\frac{2}{z-\bar z}(\partial_z-\partial_{\bar z})\,,\qquad D^+_0=\partial_z+\partial_{\bar z}\,,\qquad D^+_1=z\partial_z+\bar z\partial_{\bar z}\,,
}
which is degree 2 in $\sigma,\tau$. Since the second term in \eqref{Ward2} satisfies the Ward identities by itself, the function $\mathcal{F}_{p}(U,V;\sigma,\tau)$ must satisfy the Ward identities independently. We can fix $\mathcal{F}_{p}(U,V;\sigma,\tau)$ and $ \mathcal{H}_{p}(U,V;\sigma,\tau)$ by demanding that when we twist the four point function as
\es{twist}{
\mathcal{G}_{2p}\big\vert_{2d}( z)&\equiv \mathcal{G}_{2p}(z\bar z, (1-z)(1-\bar z),\bar z^{-2},(1-\bar z^{-1})^{2})\\
&=\mathcal{F}_{p}(z\bar z, (1-z)(1-\bar z),\bar z^{-2},(1-\bar z^{-1})^{2})\,,
}
then $ \mathcal{H}_{p}(U,V;\sigma,\tau)$ vanishes and $\mathcal{F}_{p}(U,V;\sigma,\tau)$ only depends on $z$. This definition fixes $\mathcal{F}_{p}(U,V;\sigma,\tau)$ to resemble a free theory correlator
\es{F22pp}{
\mathcal{F}_{p}(U,V;\sigma,\tau)=&1+\delta_{2,p}(\sigma U^2+\tau^2{U^4}/{V^4})+\frac{2p}{c}\left((p-1)\frac{\sigma\tau U^4}{V^2}+{\sigma U^2}+\frac{\tau U^2}{V^2}\right)\,,\\
} 
so that the twisted four-point function is
\es{Ftwist}{
\mathcal{G}_{2p}\big\vert_{2d}( z)=&1+\delta_{2,p}z^4(1+(1-z)^{-4})+\frac1c \frac{2p z^2 (z (p z-2)+2)}{(z-1)^2}\,,
}
where note that the $\bar z$-dependence has vanished. The existence of this solution to the Ward identities was first pointed out for $\langle pppp\rangle$ in \cite{Dolan:2004mu}, and in \cite{Beem:2014kka} was used to argue that 6d $(2,0)$ CFTs contain a protected 2d chiral algebra subsector that only includes the BPS multiplets $\mathcal{D}[a\,0]$ and $\mathcal{B}[b\,0]$,\footnote{The multiplets $\mathcal{D}[a,2]$ also appear in the 2d chiral algebra sector, but these do not appear in the $s$-channel of $\langle22pp\rangle$ so we do not consider them.} which can be used to compute their OPE coefficients. As we see, for $\langle22pp\rangle$ the 2d chiral algebra is $1/c$ exact,\footnote{This follows from the fact that crossing symmetry in the 2d chiral algebra fixes the correlator completely in terms of the stress tensor block, whose coefficient is fixed by conformal Ward identities to be $1/c$.} and so gives the same OPE coefficients in terms of $c$ for any 6d $(2,0)$ CFT.\footnote{For $\langle pqrs\rangle$ with $p,q,r,s>2$, the 2d chiral algebra has nontrivial $c$-dependence that depends on the physical theory, see for instance \cite{Chester:2018dga} for a detailed description of $\langle pppp\rangle$ for the $A_{N-1}$ theory, where the 2d chiral algebra is given by the $W_\infty$ algebra as conjectured in \cite{Beem:2014kka}.}

Since $\mathcal{F}_{p}(U,V,\sigma,\tau)$ is $1/c$ exact, all the nontrivial information in the theory is given by the reduced correlator $\mathcal{H}_{p}(U,V)$. To perform the superblock expansion of $\mathcal{H}_{p}(U,V)$, we must rewrite the formal solution \eqref{Ward2} as \cite{Dolan:2004mu,Beem:2015aoa}\footnote{Note that $\Upsilon$ is defined slightly differently in \cite{Rastelli:2017ymc} versus \cite{Beem:2015aoa}, which considered the case $\langle 2222\rangle$. The resulting $\mathcal{T}_{p}$'s (or similarly $\mathcal{H}_{p}$'s) are related by a factor of $U^5$. We use the definition in \cite{Rastelli:2017ymc}. }
\es{Asecond}{
\mathcal{G}_{2p}(U,V,\sigma,\tau)= U^2D\left[\frac{(z\alpha-1)(z\bar \alpha-1)h_{p}(z)-(\bar z\alpha-1)(\bar z\bar \alpha-1)h_{p}(\bar z)}{z-\bar z}\right]+\Upsilon\circ\mathcal{T}_{p}(U,V)\,,
}
so that we can now expand $\mathcal{T}_{p}(U,V)$ in reduced correlator superblocks as \cite{Beem:2015aoa}
\es{blockExpT}{
\mathcal{T}_{p}(U,V)=&\sum_{\mathcal{M}_{\Delta,\ell}\in\{\cL[00]_{\Delta,\ell},\cD[40],\cD[04],\cB[20]_\ell,\cB[02]_\ell\}} \lambda_{22\mathcal{M}} \lambda_{pp\mathcal{M}} C_{\mathcal{M}}\frac{G_{\Delta'+4,\ell'}^{0,-2}(U,V)}{U}\,,\\ 
}
where $G_{\Delta+4,j}^{0,-2}(U,V)$ is a 6d conformal block that would appear in a four-point function of operators with unequal scaling dimensions, as defined in Appendix \ref{block6dApp}. All $\mathcal{M}_{\Delta,\ell}$ in $S_2\times S_2$ except for $\cD[20]$ and $\cD[00]$ have an expansion in $\cT_p(U,V)$, and the effective dimensions $\Delta'$ and spins $\ell'$ that appear are related to $\Delta,\ell$ as described in Table \ref{moreMult}. The coefficients $C_{\mathcal{M}}$ are defined so that the pair of OPE coefficients $ \lambda_{22\mathcal{M}} \lambda_{pp\mathcal{M}} $ has the same normalization as in \eqref{SBDecomp}, and take the form\footnote{The coefficients $C_\cM$, as well as the effective dimensions and spins for each $\cM$, are related to the unique super-descendent of each $\cM$ that appears in the highest $\mathfrak{so}(5)_R$ channel.}
  \es{C}{
 C_{\mathcal{D}[00]}=&-1\,,\qquad  C_{\mathcal{D}[20]}=2\,,\qquad  C_{\mathcal{D}[40]}=2\,,\qquad  C_{\mathcal{D}[04]}=\frac13\,,\qquad  C_{\mathcal{B}[20]_\ell}=\frac{2 (\ell+3)}{( \ell+5) ( \ell+1)} \,,\\
  C_{\mathcal{B}[02]_\ell}=&\frac{ (\ell+2) }{(\ell+1) (\ell+4) }\,,\qquad C_{\mathcal{L}[00]_{\Delta,\ell}}=-\frac{4}{(-\Delta +\ell+6) (\Delta +\ell-2)}\,,
 }
 where $C_{\mathcal{D}[00]}$ and $C_{\mathcal{D}[20]}$ do not appear in \eqref{blockExpT} but will be useful below.

 \begin{table}[htp]
\begin{center}
 \begin{tabular}{c|c|c|c|c|r}
$\cM_{\Delta,\ell}$&   $\Delta$ in $\mathcal{G}_{2p}$&  $\ell$ in $\mathcal{G}_{2p}$ &$\Delta'$ in $\mathcal{T}_{p}$   & $\ell'$ in $\mathcal{T}_{p}$  & $\mathfrak{so}(5)$ irrep \\
  \hline
${\Blue\mathcal{D}{[00]}}$&  $0$ & $0$&  N/A & N/A & ${\bf 1} = [00]$ \\
${\Blue\mathcal{D}{[20]}}$&  $4$ & $0$ &  N/A & N/A& ${\bf 14} = [20]$ \\
${\Blue\mathcal{D}{[40]}}$&  $8$ & $0$ &  $4$ & $0$& ${\bf 55} = [40]$ \\
$\mathcal{D}{[04]}$&  $8$ & $0$ &  $6$ & $0$& ${\bf 35'} = [04]$ \\
${\Blue\mathcal{B}{[20]}_\ell}$&  $8+\ell$ & $\ell\in\text{even}$ &  $6+\ell$ & $\ell+2$& ${\bf 14} = [20]$ \\
$\mathcal{B}{[02]}_\ell$&  $8+\ell$ & $\ell\in\text{odd}$ &  $7+\ell$ & $\ell+1$& ${\bf 10} = [02]$ \\
$\mathcal{L}{[00]}_{\Delta,\ell}$&  $\Delta>6+\ell$ &  $\ell\in\text{even}$ & $\Delta$& $\ell\in\text{even}$ & ${\bf 1} = [00]$ \\
 \end{tabular}
\end{center}
\caption{Scaling dimensions and spins of multiplets in $S_2\times S_2$ in the full correlator $\cG_{2p}$, denoted as $(\Delta,\ell)$, versus the reduced correlator $\cT_{p}$, denoted as $(\Delta',\ell')$. The operators in the 2d chiral algebra sector are denoted in blue.}
\label{moreMult}
\end{table}%

The operators $\cD[00]$, $\cD[20]$, $\cD[40]$, and $\cB[20]_\ell$ that appear in the 2d chiral algebra must also be simultaneously expanded in $h_{p}(z)$ as
\es{hexp}{
h_{p}(z)=&-\hspace{-.2in}\sum_{\cM_{\Delta,\ell}\in \{\cD[00],\cD[20],\cD[40]\}} \hspace{-.2in}\lambda_{22\mathcal{M}} \lambda_{pp\mathcal{M}} C_{\mathcal{M}} z^{\frac{\Delta-\ell}{2}-1}g_{\Delta,\ell}^{0,-2}(1-z)\\
&-\sum_{\cM_{\Delta,\ell}\in \{\cB[20]_\ell\}} \hspace{-.2in}\lambda_{22\mathcal{M}} \lambda_{pp\mathcal{M}} C_{\mathcal{M}} z^{\frac{\Delta-\ell}{2}-1}g_{\Delta+2,\ell+2}^{0,-2}(1-z)\,,
}
where the coefficients $C_{\mathcal{M}}$ are the same as \eqref{C}, and the leading order lightcone blocks $g_{\Delta,\ell}^{\Delta_{12},\Delta_{34}}(V)$ are defined in Appendix \ref{block6dApp} and are what naturally appear in the 2d chiral algebra.

Finally, we can relate $\cT_{p}(U,V)$ to $\cH_{p}(U,V)$ by writing $\mathcal{F}_{p}$ in terms of $h_{p}(z)$ and $\cT_{p}(U,V)$ using \eqref{Ward2} and \eqref{Asecond} to get
\es{firstTosecond}{
\mathcal{F}_{p}\big\vert_{2d}( z)=&-z^2\partial_zh_{p}(z)\,,\\
\mathcal{H}_{p}(U,V)=&\mathcal{T}_{p}(U,V)+F_{p}^{(0)}(U,V)+c^{-1}F_{p}^R(U,V)\,,\\
}
where we define the functions
\es{FF}{
F_{p}^{(0)}(U,V)=&-U^3\frac{h_{p}(\bar z)\vert_{c^0}-h_{p}(z)\vert_{c^0}}{(z-\bar z)^3}+\frac{ \delta_{p,2}U^4\left(V^{-2}+1\right)+U^2}{  
   \left(U^2-2 U (V+1)+(V-1)^2\right)}\,,\\
   F_{p}^R(U,V)=&-U^3\frac{h_{p}(\bar z)\vert_{c^{-1}}-h_{p}(z)\vert_{c^{-1}}}{(z-\bar z)^3}+\frac{ {2p} U^3 (U(p-1)+V+1)}{  V
   \left(U^2-2 U (V+1)+(V-1)^2\right)}\,.
}
The block expansion of $\cH_{p}(U,V)$ then follows from that of $\mathcal{T}_{p}(U,V)$ discussed above. When $p=2$, $\cH\equiv\cH_{2}$ and $\cF\equiv\cF_{2}$ transform under the crossing $1\leftrightarrow3$ as
\es{crossing}{
\mathcal{H}(U,V)=\frac{U^4}{V^4}\mathcal{H}(V,U)\,,\qquad \mathcal{F}(U,V)=\frac{U^4}{V^4}\mathcal{F}(V,U)\,,
}
which we can use to derive crossing for $\cT(U,V)\equiv\mathcal{T}_2(U,V)$ and $h(z)\equiv h_2(z)$: 
\es{crossingT}{
\mathcal{T}(U,V)+U^3\frac{h(\bar z)-h(z)}{(z-\bar z)^3}=\frac{U^4}{V^4}\left(\mathcal{T}(V,U)-V^3\frac{h(1-\bar z)-h(1-z)}{(z-\bar z)^3}\right)\,,
}
where note that the last term in \eqref{firstTosecond} was already crossing symmetric for $p=2$, and so does not appear in the crossing equations.

\section{$A_{N-1}$ and $D_N$ correlators at large $N$}
\label{largecc}

We will now restrict to the physical $A_{N-1}$ and $D_N$ theories and discuss the large $c\sim N^3\to\infty$ limit. We begin by discussing the strict $c\to\infty$ limit, where both theories are described by the same generalized free field theory (GFFT), which we use to compute the average OPE coefficients of double-trace $\cL[00]$ multiplets in $\langle qqpp\rangle$. We then consider the $1/c$ corrections to $\langle 22pp\rangle$, which we use to compute the average anomalous dimension of the $\cL[00]$ at orders $1/c$ and $1/c^{5/3}$, which correspond to tree level supergravity and $R^4$, respectively. The $1/c$ terms will be the same for both $A_{N-1}$ and $D_N$, up to the vanishing of odd $p$ terms for $D_N$, while the $1/c^{5/3}$ values will differ by a power of the orbifold factor $\mathfrak{o}$ in \eqref{cPlanck}. Finally, we discuss the large $c$ expansion to higher orders for $\langle2222\rangle$, which will be our main focus in the rest of the paper.

\subsection{Generalized free field theory at $c\to\infty$}
\label{strong}

We start by discussing the strict $c\sim N^3\to\infty$ expansion of $\langle qqpp\rangle$ and $\langle pppp\rangle$ for the $A_{N-1}$ and $D_N$ theories. Recall from the Introduction that both theories have the same half-BPS correlators at this order, except that all correlators involving $S_p$ for odd $p$ vanish for the $D_N$ theory. In particular, both theories are described by a GFFT where the operators $S_p$ are treated as generalized free fields with two point functions $\langle S_p(x_1,Y_1) S_q(x_2,Y_2)\rangle=\delta_{pq}\frac{(Y_1\cdot Y_2)^p}{|x_{12}|^{4p}}$. We can then compute $\langle qqpp\rangle$ (for $q\leq p$) using Wick contractions to get
\es{Ninf}{
\cG_{qp}^{(0)}=1+\delta_{qp}\left(U^{2p}\sigma^p+\frac{U^{2p}}{V^{2p}}\tau^p\right)\,.
}
We can use this expression and the superblock expansions of the previous section to compute the OPE coefficients of all operators in $S_q\times S_q$. If several operators have the same quantum numbers at this order, then we can only compute the average of their OPE coefficients. Such a degeneracy occurs for double trace $\mathcal{L}[00]_{\Delta,\ell}$ operators $S_p\partial_{\mu_1}\dots\partial_{\mu_\ell}(\partial^2)^nS_p$ which have spin $\ell$ and twist $ t\equiv \Delta-\ell=4p+2n$. For $t\geq8$, there are $\lfloor{t/4-1\rfloor}$ such degenerate operators due to the different ways of adding $p$ and $n$ to get the same twist, which we label using the degeneracy label $I$. Let us denote the GFFT OPE coefficient of these operators in the $S_p\times S_p$ OPE by $\lambda^{(0)}_{p,t,\ell,I}$. From the GFFT result $ \cG_{qp}=1$ for $q\neq p$, we can immediately see that the following average vanishes:
\es{easyAv}{
\langle \lambda^{(0)}_{q,t,\ell} \lambda^{(0)}_{p,t,\ell} \rangle\equiv \sum_{I=1}^{\lfloor{t/4-1}\rfloor} \lambda^{(0)}_{q,t,\ell,I} \lambda^{(0)}_{p,t,\ell,I} =0\,,\qquad\qquad \text{for}\;\;q\neq p\,,
}
which is in fact true for all such holographic theories in the strict large $c$ limit. It is much more nontrivial to compute the average $\langle (\lambda^{(0)}_{p,t,\ell})^2\rangle\equiv \sum_I (\lambda^{(0)}_{p,t,\ell,I})^2$ that appears in $\langle pppp\rangle$. By equating the GFFT correlator $\cG_{pp}^{(0)}$ in \eqref{Ninf} to the superblock expansion in \eqref{SBDecomp}, we can compute the OPE coefficient squared of all multiplets in $S_p\times S_p$ to leading order. This is a conceptually straightforward but technically arduous calculation, since it requires the explicit superconformal blocks for each $\langle pppp\rangle$. We found it useful to use the radial expansion of conformal blocks in terms of $r,\eta$ coordinates, as described in Appendix \ref{block6dApp}, which organizes the expansion in terms of the scaling dimension of blocks. From explicitly computing the OPE coefficients for $p=2,\dots, 11$, we found the general formula
\es{ppppLam}{
\langle (\lambda^{(0)}_{p,t,\ell})^2\rangle&=\frac{\Gamma \left(\frac{t-4}{2}\right) \Gamma \left(\ell+\frac{t}{2}\right) \Gamma \left(\frac{1}{2} (4 p+t-2)\right) \Gamma \left(\ell+2 p+\frac{t}{2}+1\right)}{ \Gamma \left(\frac{t-1}{2}\right) \Gamma \left(\frac{1}{2} (2 \ell+t+3)\right) \Gamma \left(\frac{1}{2} (-4 p+t+2)\right) \Gamma
   \left(\ell-2 p+\frac{t}{2}+3\right)}\\
 &  \qquad \times \frac{3 \pi  (\ell+1) (\ell+2) 2^{-2 (\ell+t-4)} (\ell+t+1) (\ell+t+2)}{\Gamma (2
   p)^2 \Gamma (2 p-2) \Gamma (2 p+1) (t+2) (t+4) (2 \ell+t+6) (2 \ell+t+8)}\,,
}
which generalizes the $p=2$ formula previously derived from $\langle 2222\rangle$ in \cite{Heslop:2004du}. 

\subsection{Tree level $\langle 22pp\rangle$}
\label{22pptree}

Next, we further restrict to $\langle22pp\rangle$ and consider the large $c$ expansion to subleading orders  $1/c$ and $1/c^{5/3}$, which corresponds to tree level supergravity and $R^4$ in the bulk description. We can expand $\cH_{p}$ in \eqref{Ward2} to this order as
\es{Hlarge}{
\cH_{p}(U,V)=\cH^{(0)}_{p}+c^{-1}\cH^R_{p}+c^{-\frac53}\cH^{R^4}_{p}+\dots\,.
}
We can similarly expand the double-trace unprotected scaling dimensions and OPE coefficients squared as
\es{anomA}{
\Delta_{t,\ell,I}&=t+\ell+c^{-1}\gamma^{R}_{t,\ell,I}+c^{-\frac53}\gamma^{R^4}_{t,\ell,I}+\dots\,,\\
(\lambda_{p,t,\ell,I})^2&=(\lambda^{(0)}_{p,t,\ell,I})^2+c^{-1}(\lambda^{R}_{p,t,\ell,I})^2+c^{-\frac53}(\lambda^{R^4}_{p,t,\ell,I})^2+\dots\,,\\
}
where note that pairs of OPE coefficients are what naturally occur in four-point functions. A similar expansion exists for the OPE coefficients of the protected operators, although of course their scaling dimensions are fixed. Using these expansions, we can write the superblock expansion for $\cH_{p}$ at large $c$, which follows from the relation \eqref{firstTosecond} for $\cT_p$ and the block expansion of $\cT_p$ in \eqref{blockExpT}. For $\cH_{p}^R$ the expansion is
\es{SGexp}{
&\cH_{p}^R(U,V)=F_p^R(U,V)+\sum_{\mathcal{M}_{\Delta,\ell}\in\{\cD[40],\cD[04],\cB[20]_\ell,\cB[02]_\ell\}} \lambda^R_{22\mathcal{M}} \lambda^R_{pp\mathcal{M}} C_{\mathcal{M}}\frac{G_{\Delta{}'+4,\ell'}^{0,-2}(U,V)}{U}\\
&\quad+\sum_{ t,\ell,I} \left[\lambda^{R}_{2,t,\ell,I} \lambda^{R}_{p,t,\ell,I}+\lambda^{(0)}_{2,t,\ell,I} \lambda^{(0)}_{p,t,\ell,I}\gamma_{t,\ell,I}^R(\partial_t^\text{no-log}+\frac12\log U)\right]\left[ C_{t,\ell}\frac{G_{t+\ell+4,\ell}^{0,-2}(U,V)}{U}\right]\,,
}
where $F_p^R(U,V)$ was defined in \eqref{FF}, and $C_{t,\ell}\equiv C_{\cL[00]_{t+\ell,\ell}}$. The notation $\partial_t^\text{no-log}G_{t+\ell+4,\ell}^{0,-2}$ means that we consider the term after taking the derivative that does not include a $\log U$, since the terms with a log have already been written separately. The expansion of $\cH_{p}^{R^4}$ can similarly be found by setting $R\to R^4$ and noting that $F^R_p(U,V)$ no longer contributes.

While the superblock expansion is most usefully expressed in position space, to actually compute correlators in the large $c$ limit, it is more convenient to work with the Mellin transforms $\mathfrak{M}_p(s,t;\sigma,\tau)$ and $M_p(s,t)$ of the connected full and reduced correlators $\mathcal{G}^\text{con}_{2p}(U,V;\sigma,\tau)\equiv\mathcal{G}_{2p}(U,V;\sigma,\tau)-\mathcal{G}^{(0)}_{2p}(U,V;\sigma,\tau)$ and $\cH_p(U,V)$, respectively, which are defined as \cite{Rastelli:2017ymc}:
\es{mellinH}{
\mathcal{G}^\text{con}_{2p}(U,V;\sigma,\tau)&=\int\frac{ds\, dt}{(4\pi i)^2} U^{\frac s2}V^{\frac t2-p-2}\mathfrak{M}_p(s,t;\sigma,\tau) \Gamma\left[2p-\frac s2\right]\Gamma\left[4-\frac s2\right]\Gamma^2\left[p+2-\frac t2\right]\Gamma^2\left[p-1-\frac {{u}}{2}\right],\\
\mathcal{H}_{p}(U,V)&=\int\frac{ds\, dt}{(4\pi i)^2} U^{\frac s2+1}V^{\frac t2-p-1}{M}_p(s,t)\Gamma\left[2p-\frac s2\right]\Gamma\left[4-\frac s2\right]\Gamma^2\left[p+2-\frac t2\right]\Gamma^2\left[p+2-\frac {{u}}{2}\right]\,,
}
where $u = 4p+2 - s - t$ and the integration contours include all poles of the Gamma functions on one side or the other of the contour. The Mellin transform $\mathfrak{M}_p(s,t;\sigma,\tau)$ of the full correlator is defined such that a bulk contact Witten diagram coming from a vertex with $2m$ derivatives gives rise to a polynomial $\mathfrak{M}_p(s,t;\sigma,\tau)$ in $s,t$ of degree $m$, and similarly an exchange Witten diagrams corresponds to $\mathfrak{M}_p(s,t;\sigma,\tau)$ with poles for the twists of each exchanged operator. The reduced correlator Mellin amplitude $M_p(s,t)$ is then related to $\mathfrak{M}_p(s,t;\sigma,\tau)$ by the Mellin space version of  $\Upsilon$ in \eqref{Y}, which takes the form of a difference operator given in \cite{Rastelli:2017ymc} whose explicit form we will not use. The degree of a given term in $M_p(s,t)$ is seven less than that of $\mathfrak{M}_p(s,t;\sigma,\tau)$ in the large $s,t$ limit due to this difference operator. These requirements  should be supplemented by the crossing symmetry relations
 \es{crossM}{
  M_p(s,t) = M_p(s,u) \,,\qquad M_2(s,t) = M_2(t,s)\,,
 }
which follow from interchanging the first and second operators, and, for $p = 2$, the first and third.  All these requirements, namely the analytic structure, growth at infinity, and crossing symmetry of $M_p(s, t)$ imply that $M_p(s, t)$ can be expanded similar to the position space expression \eqref{Hlarge} to get
\es{Mplarge}{
M_{p}(s,t)=c^{-1}B^R(p)M^R_{p}+c^{-\frac53}B^{R^4}(p)M^{R^4}_{p}+\dots\,,
}
with 
\es{Ms}{
M^R_p=&\frac{1 }{(s-6)(s-4)(t-2p-2)(t-2p)( u-2p-2)( u-2p)}\,,\\
M^{R^4}_p=&\frac{1 }{(s-6)(t-2p-2)( u-2p-2)}\,,\\
}
where $M_p^R$ appeared previously in \cite{Rastelli:2017ymc}. From \eqref{mellinH} we can then write down the analogous position space expressions in \eqref{Hlarge} as
\es{Hs}{
\cH^R_p=&B^R(p)\frac{U^{2p+1}}{64V}\bar D_{2p+3,2p-1,3,3}(U,V)\,,\\
\cH^{R^4}_p=&-B^{R^4}(p)\frac{  V U^{ 2 p+1}}{8}\bar D_{2 (p+1),2
   (p+1),6,4}(U,V)\,,
}
where explicit expressions for $\bar D_{r_1,r_2,r_3,r_4}(U,V)$ \cite{Eden:2000bk} can be derived from e.g.~Appendix C in \cite{Binder:2018yvd}, and here we include the explicit coefficients $B^R(p)$ and $B^{R^4}(p)$ to match \eqref{Hlarge}. The coefficient $B^R(p)$ can be found by demanding that no unphysical twist 4 unprotected operators appear in the superblock expansion \eqref{blockExpT}. In our normalization we find 
\es{BSG}{
B^R(p)=\frac{64p}{\Gamma(2p-2)}\,.
}
Note that this coefficients is the same for both $A_{N-1}$ and $D_N$, since neither theory has twist 4 long operators at large $c$. 
 
 The coefficient $B^{R^4}(p)$, on the other hand, will differ for $A_{N-1}$ and $D_N$. This coefficient can be fixed using the flat space limit formula \cite{Penedones:2010ue,Chester:2018dga}, which relates Mellin amplitudes $M^a_p(s,t)$ of large $s,t$ degree $a-7$ (i.e. degree $a$ in $\mathfrak{M}_p(s,t)$) to the 11d amplitude defined in \eqref{A} as
 \es{flat}{
c^{\frac{2(1-a)}{9}} \frac{ \Gamma (2 p) \Gamma (2 p-2)}{2^{a+4}\Gamma (a+2 p)} \lim_{s,t\to\infty} (stu)^2M^a(s,t)= \ell_{11}^{2a-2}\frac{{\mathcal{A}_{2a+7}}}{\cA_R}\,,
 }
 where $\mathcal{A}_{2a+7}$ is a term in the amplitude with length dimension $(2a+7)$, and $\ell_{11}$ is the 11d Planck length. For instance, $\mathcal{A}_{15}\equiv\mathcal{A}_{R^4}$ has length dimension 15 in \eqref{A} and corresponds to $M_p^4\equiv M_p^{R^4}$. The 11d amplitude of course is the same for all $p$, and the ratio of $\mathcal{A}_{R^4}/\mathcal{A}_R$ was given in \eqref{SGtoR4}. Using this 11d amplitude and the flat space limit we find 
 \es{BR4}{
B^{R^4}(p)=\frac{2 p (p+1) (2 p+1) (2 p+3)}{3 (2\mathfrak{o})^{2/3} \Gamma (2 p-2)}\,,
}
 where $\mathfrak{o}$ is one/two for $A_{N-1}$/$D_N$. This factor comes from the different relation between $\ell_{11}$ and $c$ in \eqref{cPlanck} for each theory, and is in fact the only difference between the $R^4$ correlators of each theory. 
 
Using the explicit tree level amplitudes in \eqref{Hs} with coefficients \eqref{BSG} and \eqref{BR4}, we can expand in superblocks \eqref{SGexp} to extract the average anomalous dimensions $\langle \lambda^{(0)}_{2,t,\ell} \lambda^{(0)}_{p,t,\ell}\gamma_{t,\ell}\rangle\equiv\sum_{ I} \lambda^{(0)}_{2,t,\ell,I} \lambda^{(0)}_{p,t,\ell,I}\gamma_{t,\ell,I}$ weighted by OPE coefficients. For supergravity we find
\es{averageSG}{
\langle \lambda^{(0)}_{2,t,\ell} \lambda^{(0)}_{p,t,\ell}\gamma^R_{t,\ell}\rangle=&-\frac{\pi  (t-6)(2\ell+t-2) (4 (\ell+2) p (\ell+t+1)+(t-2) t)}{ 4^{ (\ell+t+1)}  (\ell+3) (\ell+t)} \\
&\qquad\times\frac{\Gamma \left(\frac{t}{2}+1\right) \Gamma \left(\ell+\frac{t}{2}+3\right) \Gamma \left(2 p+\frac{t}{2}-1\right)}{\Gamma (2 p) \Gamma (2 p-2) \Gamma
   \left(\frac{t-1}{2}\right) \Gamma \left(\ell+\frac{t}{2}+\frac{3}{2}\right) \Gamma \left(-2 p+\frac{t}{2}+1\right)}\,,
}
which is identical for both $A_{N-1}$ and $D_N$. For $R^4$ we find
\es{averageR4}{
\langle \lambda^{(0)}_{2,t,\ell} \lambda^{(0)}_{p,t,\ell}\gamma^{R^4}_{t,\ell}\rangle=&
\delta_{\ell,0}\frac{\pi  t (6-t) (t^2-16)  \left(t^2-4\right)^2 \Gamma \left(\frac{t}{2}+1\right) \Gamma \left(\frac{t}{2}+4\right) \Gamma \left(2
   p+\frac{t}{2}+1\right)}{3(2\mathfrak{o})^{\frac23} 4^{t+7} \Gamma (2 p) \Gamma (2 p-2) \Gamma \left(\frac{t+1}{2}\right) \Gamma \left(\frac{t+5}{2}\right) \Gamma \left(-2 p+\frac{t}{2}+1\right)}
\,,
}
which is only nonzero for zero spin and differs for each theory by the orbifold factor $\mathfrak{o}$.

\subsection{Large $c$ expansion of $\langle2222\rangle$}
\label{2222largec}

Finally, we restrict further to $\langle2222\rangle$, which is the correlator we are primarily interested in studying. To reduce clutter, we will drop the $p=2$ subscript from all expressions. The Mellin amplitude $M(s,t)$ is fixed by the analytic structure, growth at infinity, and crossing symmetry to take the form
\es{M2222}{
M(s,t)=&c^{-1}B^RM^R+c^{-\frac53}B^{R^4}_4M^4+c^{-2}(M^{R|R}+B^{R|R}_4M^4)\\
&+c^{-\frac73}(B^{D^6R^4}_4M^4+B^{D^6R^4}_6M^6+B^{D^6R^4}_7M^7)\\
&+c^{-\frac{23}{9}}(B^{D^8R^4}_4M^4+B^{D^8R^4}_6M^6+B^{D^8R^4}_7M^7+B^{D^8R^4}_8M^8)\\
&+c^{-\frac83}(M^{R|R^4}+B^{R|R^4}_4M^4+B^{R|R^4}_6M^6+B^{R|R^4}_7M^7+B^{R|R^4}_8M^8)+\dots\,,
}
where the coefficient of each $c^{-b}$ must include all allowed Mellin amplitudes of large $s,t$ degree $(9/2b-21/2)$ or less. These can include contact Mellin amplitudes $M^a$ of degree $a-7$, which take the form \cite{Chester:2018dga}
\es{contactM}{
M^a=\frac{(s^2+t^2+u^2)^{a_1}(stu)^{a_2}}{(s-6)(t-6)(u-6)}\qquad \text{s.t.}\qquad 2a_1+3a_2\leq a-4\,.
}
For the amplitudes we consider above, this implies one allowed contact amplitudes of each degree. In the full correlator Mellin amplitude $\mathfrak{M}$, these amplitudes would be polynomials of degree $a$, which correspond to contact Witten diagrams with $2a$ derivatives. These contact diagrams only contribute to a finite number of spins that grows with the degree \cite{Heemskerk:2009pn}. For the multiplets in $\langle2222\rangle$, the contribution from each $M^a$ in \eqref{M2222} is summarized in Table 2 of \cite{Chester:2018dga}, which we repeat here in Table \ref{resultList}.\footnote{We correct some typos for the anomalous dimensions.} The other Mellin amplitudes shown in \eqref{M2222} include the tree level supergravity term $M^R$ discussed in the previous section, which includes poles for the single trace supergravity multiplet, as well as the 1-loop Mellin amplitudes $M^{R|R}$ and $M^{{R^4}|R}$ of degrees $-\frac32$ and $\frac{3}{2}$, respectively, that we will discuss more in the following section. 

\begin{table}[htp]
\begin{center}
\begin {tabular} {| c || c | c | c | c | }
\hline
 {CFT data:}&$M^{4}$  &$M^{6}$ & $M^{7}$ &$M^{8}$  \\
  \hline
\TBstrut  $\lambda^2_{\mathcal{D}[04]}$&  $-\frac{3}{7}$& $-\frac{1308}{77}$ & $-\frac{1224}{77}$ & $-\frac{692304}{1001}$ \\
  \hline
 \TBstrut  $\lambda^2_{\mathcal{B}[02]_1}$& 0& $-\frac{2000}{1573}$ & $\frac{6000}{1573}$ & $-\frac{115200}{1573}$ \\
  \hline
 \TBstrut  $\lambda^2_{\mathcal{B}[02]_3}$& 0& $0$ & $0$ & $-\frac{1354752}{158015}$ \\
  \hline
\TBstrut  $\lambda^2_{\mathcal{B}[02]_5}$& 0& $0$ & $0$ & $0$ \\
  \hline
 \TBstrut $\g_{{8,0}}$& $-\frac{360}{11}$ & $-\frac{283680}{143}$ & $-\frac{138240}{143}$ & $-\frac{17493120}{143}$ \\
  \hline
 \TBstrut   $\g_{{8,2}}$& 0& $-\frac{21600}{143}$ & $\frac{86400}{143}$ & $-\frac{35078400}{2431}$ \\
  \hline
 \TBstrut     $\g_{{8,4}}$& 0& $0$& $0$ & $-\frac{11612160}{4199}$\\
  \hline
 \TBstrut    $\g_{{8,6}}$& 0& $0$& $0$ & $0$ \\
  \hline
\end{tabular}
\end{center}
\caption{Contributions to the OPE coefficients squared $\lambda^2_{22\cM}$ of some short multiplets $\cD[04]$ and $\cB[02]_\ell$ for odd $\ell$ that are not in the 2d chiral algebra sector, as well as to the anomalous dimensions $\g_{8,\ell}$ for even $\ell$ of the lowest twist $t=8$ unprotected multiplet $\cL[00]$, from large $s,t$ degree $a-7$ contact Mellin amplitudes $M^{a}(s,t)$. Adapted from \cite{Chester:2018dga}, with typos fixed.
}\label{resultList}
\end{table}

The coefficients $B^R\equiv B^R(2)$ and $B^{R^4}_4\equiv B^{R^4}(2)$ were fixed in the previous section, but the other coefficients remain unknown. As such, we can only extract CFT data up to order $c^{-\frac53}$. For the 2d chiral algebra multiplets $\cD[00]$, $\cD[20]$, $\cD[40]$, and $\cB[20]_\ell$, their OPE coefficients are in fact $1/c$ exact as discussed and in our conventions are:
\es{treedata}{
&\lambda^2_{\mathcal{D}[00]}=1,\qquad \lambda^2_{\mathcal{D}[20]}=4 c^{-1}\,,\qquad \lambda^2_{\mathcal{D}[40]}=\frac13+\frac{44}{30} c^{-1}\,,\\
& \lambda^2_{\mathcal{B}[20]_\ell}=\frac{\sqrt{\pi }  (\ell+1) (\ell+4) (\ell+5) \Gamma
   (\ell+9)}{9 \cdot4^{\ell+6}\Gamma \left(\ell+\frac{11}{2}\right)}+c^{-1}\frac{\sqrt{\pi } (\ell+1) (\ell (\ell+11)+29) \Gamma (\ell+6)}{2^{2 \ell+7}  (\ell+3) \Gamma
   \left(\ell+\frac{11}{2}\right)}\,.\\
}

For the protected operators $\cD[04]$, and $\cB[02]_\ell$ that are not in the 2d chiral algebra, their OPE coefficients receive corrections at all orders in $1/c$. The $\cB[02]_\ell$ has odd $\ell$, so its OPE coefficient does not receive a correction from the $R^4$ contact term, i.e. $c^{-\frac53}$, which only contributes to zero spin. We can thus only compute its value up to $1/c$  \cite{Beem:2015aoa,Heslop:2017sco}:
\es{treedata2}{
 \lambda^2_{\mathcal{B}[02]_\ell}&=\frac{(\ell+1) (\ell+3) (\ell+4) (\ell+8) (\ell+9) \Gamma (\ell+6) \Gamma (\ell+7)}{36 \Gamma (2 \ell+11)}\\
&-c^{-1}\frac{40 (\ell+1) (\ell (\ell+11)+27) \Gamma (\ell+6) \Gamma (\ell+7)}{ (\ell+2) (\ell+7) \Gamma (2 \ell+11)}+O(c^{-2})\,,\\
}
which at this order is the same for the $A_{N-1}$ and $D_N$ theories. The $\cD[04]$, however, has zero spin, so its OPE coefficient receives an $R^4$ correction that is different for the  $A_{N-1}$ and $D_N$ theories by the usual orbifold factor $\mathfrak{o}$ \cite{Heslop:2004du,Chester:2018dga}:
\es{D04}{
\lambda^2_{\mathcal{D}[04]}&=\frac23 - c^{-1}\frac{170}{21 }-c^{-\frac53}\frac{{60}}{(2\mathfrak{o})^{\frac23}}+O(c^{-2})\,.  
}
Finally, the average anomalous dimensions weighted by OPE coefficients for the long multiplets was already given in \eqref{averageSG} and \eqref{averageR4}. For the lowest two twists, there is no degeneracy, so we can divide these expressions by the GFFT OPE coefficients in \eqref{ppppLam} for $p=2$ to get
\es{anomLowTwist}{
\Delta_{8,\ell}=&8-c^{-1}\frac{17280}{(\ell+9) (\ell+10) \left(\ell^2+3 \ell+2\right)}-c^{-\frac53}\delta_{\ell,0}\frac{50400 }{11 (2\mathfrak{o})^{\frac23}}+O(c^{-2})\,,\\
\Delta_{10,\ell}=&10-c^{-1}\frac{120960 (\ell (\ell+13)+32)}{ (\ell+1) (\ell+2) (\ell+3) (\ell+10) (\ell+11) (\ell+12)}-c^{-\frac53}\delta_{\ell,0}\frac{12700800 }{143(2\mathfrak{o})^{\frac23}}+O(c^{-2})\,,\\
}
where as usual the difference between the $A_{N-1}$ and $D_N$ theories enter only in the $c^{-\frac53}$ correction. In the following section, we will determine the 1-loop corrections to some of this non-trivial CFT data.

\section{$\langle2222\rangle$ at 1-loop}
\label{1loop}

We now discuss the 1-loop terms in the large $c\sim N^3$ expansion of $\langle2222\rangle$. In particular, we focus on $R|R$ at $c^{-2}$, $R|R^4$ at $c^{-\frac83}$, and $R^4|R^4$ at $c^{-\frac{10}{3}}$. For each 1-loop term in both the $A_{N-1}$ and $D_N$ theories, we derive the double-discontinuity (DD) from tree and GFFT data, and then use it to write the entire correlator in Mellin space using crossing symmetry up to contact term ambiguities. We then compare the correlators for both theories to the relevant 1-loop corrections to the 11d S-matrix in the flat space limit, and find a precise match for all 1-loop amplitudes in both $A_{N-1}$ and $D_N$. Finally, we extract all low-lying CFT data using two methods: an inversion integral applied to the DD that extracts the data in terms of a compact integral expression that does not apply to certain low values of spin \cite{Alday:2016njk,Caron-Huot:2017vep}, and a projection method applied to the entire Mellin amplitude that works for all spins (up to the contact term ambiguities that affect the low spins), but is less compact \cite{Heemskerk:2009pn,Chester:2018lbz}. Note that for the $D_N$ theory, we did not find a closed form expression for the Mellin amplitude, but nevertheless we computed enough terms to be able to check both the flat space limit and extract all CFT data up to high precision.

\subsection{One-loop from tree level}

We begin by expanding the reduced correlator $\cH$ for $\langle2222\rangle$ to 1-loop order at large $c$ using the block expansion described in section \ref{22pp}. For $R|R$ at order $c^{-2}$, this takes the form
\es{RR}{
\cH^{R|R}=&\sum_{t=8,10,\dots}\sum_{\ell\in\text{Even}}\Big[\frac18\langle(\lambda^{(0)}_{t,\ell})^2(\gamma^R_{t,\ell})^2\rangle(\log^2U+4\log U\partial_t^\text{no-log}+4(\partial_t^\text{no-log})^2) \\
&+\frac12\langle(\lambda^{R})^2_{t,\ell}\gamma^{R}_{t,\ell}\rangle(\log U+2\partial_t^\text{no-log})\\
 &+\frac12\langle(\lambda^{(0)}_{t,\ell})^2\gamma^{{R}|{R}}_{t,\ell}\rangle(\log U+2\partial_t^\text{no-log})+\langle(\lambda^{{R}|{R}}_{t,\ell})^2\rangle\Big] C_{t,\ell}U^{-1}G_{t+\ell+4,\ell}^{0,-2}(U,V)\\
 &+(\lambda_{\cD[04]}^{R|R})^2C_{\cD[04]} U^{-1}G^{0,-2}_{10,0}(U,V)+\sum_{\ell\in\text{Odd}}(\lambda_{\cB[02]_\ell}^{R|R})^2C_{\cB[02]_\ell }U^{-1}G^{0,-2}_{\ell+11,\ell+1}(U,V)\,,
}
where the normalization factors $C_{t,\ell}\equiv C_{\cL[00]_{t+\ell,\ell}}$, $C_{\cB[02]_\ell}$, and $C_{\cD[04]}$ are defined in \eqref{C}, we suppressed the $p=2$ subscripts for simplicity, and $\partial_t^\text{no-log}G_{t+\ell+4,\ell}^{0,-2}$ was defined in \eqref{SGexp}. The first three lines describe the double trace long multiplets $\cL[00]_{t+\ell,\ell}$, which are the only long multiplets that appear at this order. As described in Section \ref{largecc}, long multiplets of twist $t$ are $\lfloor t/4-1\rfloor$-fold degenerate, so only the average CFT data denoted by $\langle\rangle$ appears. The fourth line includes the protected multiplets $\cD[04]$ and $\cB[02]_\ell$, which unlike the  protected multiplets in the 2d chiral algebra sector, include $1/c$ corrections beyond $c^{-1}$. The expression for $\cH^{R^4|R^4}$ at order $c^{-\frac{10}{3}}$ is identical except we replace $R\to R^4$ and the sum for the long multiplets is now restricted to $\ell=0$. The expression for $\cH^{R|R^4}$ at order $c^{-\frac83}$ takes the similar form
\es{RR4}{
\cH^{R|R^4}=&\sum_{t=8,10,\dots}\Big[\frac14\langle(\lambda^{(0)}_{t,0})^2\gamma^R_{t,0}\gamma^{R^4}_{t,0}\rangle(\log^2U+4\log U\partial_t^\text{no-log}+4(\partial_t^\text{no-log})^2)\\
& +\frac12\langle(\lambda^{R})^2_{t,0}\gamma^{R^4}_{t,0}+(\lambda^{R^4})^2_{t,0}\gamma^{R}_{t,0}\rangle(\log U+2\partial_t^\text{no-log})\\
 &+\frac12\langle(\lambda^{(0)}_{t,0})^2\gamma^{{R}|{R^4}}_{t,0}\rangle(\log U+2\partial_t^\text{no-log})+\langle(\lambda^{{R}|{R^4}}_{t,0})^2\rangle\Big] C_{t,0}U^{-1}G_{t+4,0}^{0,-2}(U,V)\\
&+(\lambda_{\cD[04]}^{R|R^4})^2C_{\cD[04]} U^{-1}G^{0,-2}_{10,0}(U,V)+\sum_{\ell\in\text{Odd}}(\lambda_{\cB[02]_\ell}^{R|R^4})^2C_{\cB[02]_\ell }U^{-1}G^{0,-2}_{\ell+11,\ell+1}(U,V)\,,
}
where note that the first line has a factor of two relative to \eqref{RR}, and only $\ell=0$ appears due to the $R^4$.

As shown in \cite{Aharony:2016dwx}, the entire 1-loop term up to the contact term ambiguities described in Section \ref{2222largec} can in fact be constructed from the $\log^2 U$ terms shown above, which are written in terms of GFFT and tree data, since under $1\leftrightarrow3$ crossing \eqref{crossing} these are related to $\log^2V$ terms that are the only contributions to the DD at this order. The DD can then be used to reconstruct the full 1-loop correlator as shown in \cite{Alday:2016njk,Caron-Huot:2017vep}. A subtlety in the case of holographic theories is that the average $\langle(\lambda^{(0)}_{t,\ell})^2\gamma^A_{t,\ell}\gamma^{B}_{t,\ell}\rangle$ for 1-loop vertices $A,B$ is what appears in the $\log^2U$ term, whereas the different averages $\langle(\lambda^{(0)}_{t,\ell})^2\gamma^A_{t,\ell}\rangle$ and $\langle(\lambda^{(0)}_{t,\ell})^2\gamma^B_{t,\ell}\rangle$ are what appear at tree level. As shown in Appendix A of \cite{Alday:2018pdi} for the similar case of $\mathcal{N}=4$ SYM, one can compute $\langle(\lambda^{(0)}_{t,\ell})^2\gamma^A_{t,\ell}\gamma^{B}_{t,\ell}\rangle$ from GFFT $\langle ppqq\rangle$ and tree level $\langle22pp\rangle$ data as
\es{appA}{
\langle(\lambda^{(0)}_{t,\ell})^2\gamma^A_{t,\ell}\gamma^{B}_{t,\ell}\rangle=\sum_{p=2}^{\lfloor t/4\rfloor}\frac{\langle\lambda^{(0)}_{2,t,\ell}\lambda^{(0)}_{p,t,\ell}\gamma^A_{t,\ell}\rangle  \langle\lambda^{(0)}_{2,t,\ell}\lambda^{(0)}_{p,t,\ell}\gamma^B_{t,\ell}\rangle}{ {\langle(\lambda^{(0)}_{p,t,\ell})^2\rangle} }\,,
}
where we summed over each $p$ where a given twist $t$ long multiplet appears. For $A_{N-1}$, this include all integer $p$, while for $D_N$ it only includes even $p$. We computed ${\langle(\lambda^{(0)}_{p,t,\ell})^2\rangle} $, $\langle\lambda^{(0)}_{2,t,\ell}\lambda^{(0)}_{p,t,\ell}\gamma^R_{t,\ell}\rangle$, and $\langle\lambda^{(0)}_{2,t,\ell}\lambda^{(0)}_{p,t,\ell}\gamma^{R^4}_{t,\ell}\rangle$ in \eqref{ppppLam}, \eqref{averageSG}, and \eqref{averageR4}, respectively, which is sufficient to compute $R|R$, $R|R^4$, and $R^4|R^4$ for the $A_{N-1}$ and $D_N$ theories. The $p,t,\ell$ sums for the $\log^2U$ term in $R|R$ can be done by expanding at small $z$:
\es{slices}{
\frac18\sum_{t=8,10,\dots} \sum_{\ell\in\text{Even}}\sum_{p=2}^{\lfloor t/4\rfloor}\frac{\langle\lambda^{(0)}_{2,t,\ell}\lambda^{(0)}_{p,t,\ell}\gamma^R_{t,\ell}\rangle^2  }{ {\langle(\lambda^{(0)}_{p,t,\ell})^2\rangle} }C_{t,\ell}\frac{G_{t+\ell+4,\ell}^{0,-2}(U,V)}{U}=  z^5 h_{R|R}^{(5)}(\bar z) + z^6 h_{R|R}^{(6)}(\bar z) + \cdots \,,
}
and similarly for $R|R^4$ and $R^4|R^4$ using the general expression \eqref{appA}, except the $\ell$ sum is trivially $\ell=0$ in those cases, and $R|R^4$ has an extra factor of 2. The $z$-slices for $R|R$ for both $A_{N-1}$ and $D_N$ take the form
\es{slices2}{
h^{(n)}_{R|R}(\bar z) =\frac{P^{(n-1)}_{1,R|R}(\bar z)}{\bar z^{n}}\log ^2(1-\bar z) + \frac{P^{(n-2)}_{2,R|R}(\bar z)}{\bar z^{n-1}}\text{Li}_2(\bar z)+ \frac{P^{(n-2)}_{3,R|R}(\bar z)}{\bar z^{n-1}}\log(1-\bar z)  + \frac{P^{(n-2)}_{4,R|R}(\bar z)}{\bar z^{n-2}}\,,
}
where $P^{(n-1)}_{1,R|R}(\bar z), P^{(n-2)}_{2,R|R}(\bar z), P^{(n-2)}_{3,R|R}(\bar z)$ are polynomials of degree $n-1,n-2$ and $n-2$ which vanish at $\bar z=1$, and $P_{4,R|R}^{(n-2)}(\bar z)$ is a polynomial of degree $n-2$. The slices for $R|R^4$ take the simpler form
\es{slices3}{
h^{(n)}_{R|R^4}(\bar z) = \frac{P^{(n)}_{1,R|R^4}(\bar z)}{\bar z^{n+1}}\log(1-\bar z)  + \frac{P^{(n)}_{2,R|R^4}(\bar z)}{\bar z^{n}}\,,
}
where $P^{(n)}_{1,R|R^4}(\bar z)$ and $P^{(n)}_{2,R|R^4}(\bar z)$ are polynomials of degree $n$, and a similar expression holds for $R^4|R^4$. For $n=5,6$ the slices for both $A_{N-1}$ and $D_N$ theories coincide. For $n\geq7$ they are different, although their structure is identical, and we give the explicit polynomials for many $n$ in the attached \texttt{Mathematica} file. 

\subsection{Mellin amplitude}
\label{reducedMellinAmplitude}

We now show how to complete the position space DD to the entire correlator using crossing symmetry in Mellin space. For $A_{N-1}$, we will find a closed form expression for all the 1-loop Mellin amplitudes, while for the $D_N$ theory we will find closed form expressions in the large $s,t$ limit, and show how to compute as many terms as needed for the finite $s,t$ amplitude. This will be sufficient for the flat space limit and CFT extraction of the later sections.

We can compute the Mellin amplitudes from the resummed DD's following a very similar calculation in $\mathcal{N}=4$ SYM in \cite{Alday:2018kkw}. In the previous section, we computed the coefficient of $\log^2U$ in the $s$-channel, which gave the DD in the $t$-channel, as an expansion in small $z$. To convert to Mellin space, its convenient to rewrite the $z$ coefficients $z^nh^{(n)}(\bar z)$ in \eqref{slices} as $U$ coefficients $U^n \tilde h^{(n)}(V)$. For instance, for the $R|R$ expansion in \eqref{slices} (and restoring the $\log^2U$ that multiplies this), we have
\es{slices22}{
\log^2 U\left[z^5 h^{(5)}_{R|R}(\bar z) + z^6 h^{(6)}_{R|R}(\bar z) + \cdots\right]=\log^2 U\left[U^5 \tilde h^{(5)}_{R|R}(V) + U^6 \tilde h^{(6)}_{R|R}(V) + \cdots\right]\,,
}
where $\tilde h_{R|R}^{(n)}(V)$ are related to as $h_{R|R}^{(n)}(\bar z)$ as
\es{ztoU}{
\tilde h_{R|R}^{(5)}(V) = \frac{h_{R|R}^{(5)}(1-V) }{(1-V)^5}\,,\qquad \tilde h_{R|R}^{(6)}(V) = \frac{h^{(6)}(1-V) }{(1-V)^6}+ \frac{5Vh_{R|R}^{(5)}(1-V) }{(1-V)^7}+ \frac{\partial_Vh_{R|R}^{(5)}(1-V) }{(1-V)^6}\,.
}
From the definition of the Mellin transform in \eqref{mellinH}, we can then convert $U^n\log^2U\tilde h^{(n)}(V)$ to an $s$-pole in $M(s,t)$ as
\begin{equation}
 U^{n}  \log^2 U \tilde h_n(V) \leftrightarrow \frac{res_{n-1}(t)}{s-2(n-1)}\,,
\end{equation}
where the residues $res_{n-1}(t)$ follows from the $t$-integral in \eqref{mellinH}. Note that all the position space slices began at $n=5$, which correspond to the double poles in \eqref{mellinH} that give $\log^2U$ terms. We can then use crossing symmetry \eqref{crossM} to fix the other parts of the Mellin amplitude that are analytic in $s$.

We have carried out this procedure for both $A_{N-1}$ and $D_N$, and found a similar structure for both. For $M^{R|R}$ we got
\es{MellinRR}{
&M^{R|R}(s,t) = \sum_{m,n=4}^\infty \frac{c_{mn}}{(s-2m)(t-2n)}   + \sum_{m=4} \frac{c_m}{s-2m} \left(\frac{1}{t-6}+\frac{1}{u-6} \right) +\text{crossed}\,,
}
where the coefficients $c_{mn}=c_{nm}$ and $c_m$ are related for each theory as
\es{cmn}{
c^D_{mn} &= \frac{1}{2}  c^A_{mn} +  d_{mn}\,,\qquad\qquad c^D_{m} = \frac{1}{2}  c^A_{m} +  d_{m}\,.
}
The $ c^A_{mn}$ have the closed form
\es{cmn2}{
 c^A_{mn} = \frac{\Gamma (m-3) \Gamma (n-3)}{\Gamma (m+n-1)} R^{(7)}(m,n) +\frac{\Gamma (m-3) \Gamma (n-3) \Gamma \left(m+n-\frac{17}{2}\right)}{\Gamma \left(m+\frac{3}{2}\right) \Gamma \left(n+\frac{3}{2}\right) \Gamma (m+n-1)} R^{(9)}(m,n) \,.
}
For $c^A_m$, the lowest few values are
\begin{eqnarray}
\label{cmhatlow}
c^A_4=\frac{3198}{175}, ~~~c^A_5=\frac{15187}{1575},~~~ c^A_6 = \frac{7342178}{848925},~~~c^A_7 =\frac{146404341}{19619600}\,,
\end{eqnarray}
while for $m \geq 8$ we have the closed form
\es{cmhat}{
 c^A_{m} =\frac{S^{(7)}(m)}{(m-3)^2 (m-2)^2 (m-1)^2 m^2 (m+1)^2}
+ \frac{S^{(8)}(m) \Gamma (m-3)}{(m-3) (m-2) (m-1) m (m+1) \Gamma \left(m+\frac{1}{2}\right)} \,.
}
The explicit polynomials $R(m,n)$ and $S(m)$ are given in the attached \texttt{Mathematica} notebook. At large $m,n$ the $c^A_{mn}$ and $c^A_{m}$ go as 
\es{largecmn}{
 c^A_{m,n\gg1} = -\frac{3828825 \sqrt{\pi } m^{3/2} n^{3/2}}{262144 (m+n)^{9/2}}\,,\qquad\qquad  c^A_{m\gg1} =\frac{328185 \sqrt{\pi } \sqrt{\frac{1}{m}}}{32768}\,. 
}
 We have not found a closed form expression for $ d_{m,n}$ or $d_m$, but we did observe that they are subleading in the large $m,n$ limit. From the asymptotic behavior of $c^A_{mn}$ and $c^A_m$ we see that all these sums are convergent, unlike the $M^{R|R}$ in $\mathcal{N}=4$ SYM derived in \cite{Alday:2018kkw}, which had to regularized by an infinite contact term. This convergence is not so important, however, since in both 4d and 6d the full $c^{-2}$ term includes $M^{R|R}$ as well as the contact term $M^4$ as discussed in Section \ref{2222largec}.
 
 Similar results can be obtained for the reduced Mellin amplitudes $M^{R|R^4}(s,t)$ and $M^{R^4|R^4}(s,t)$ whose structure is much simpler. For $M^{R|R^4}(s,t)$ we find
\es{RR4c}{
M^{R|R^4}(s,t) &= \sum_{m=4}^\infty\left( \frac{\hat c_{m}}{(s-2m)(t-6)(u-6)} +\text{crossed}    \right)\,,
}
while $M^{R^4|R^4}(s,t)$ takes the same form except with different coefficients $\hat{\hat c}_m$. The coefficients in each case are related for the $A_{N-1}$ and $D_N$ theories as
\es{RR4cm}{
\hat c^{D}_{m} &= \frac{1}{2} \frac{\hat c^A_m}{ 2^{2/3}} + \hat d_m\,,\qquad\qquad \hat{\hat c}^{D}_{m} = \frac{1}{2} \frac{\hat {\hat c}^A_m}{(2)^{4/3}} + \hat {\hat d}_m\,,
}
where each factor of $2^{\frac23}$ comes from the orbifold factor $\mathfrak{o}$ for $R^4$. The coefficients $\hat c^A_m$ and $\hat {\hat c}^A_m$ can again be found in a closed form that we give in the attached \texttt{Mathematica} notebook. At large $m$ they go as
\es{largemRR4}{
\hat c^A_{m\gg1} = -\frac{1673196525 \sqrt{\pi } m^{7/2}}{4194304\cdot 2^{\frac23}}\,,\qquad\qquad \hat {\hat c}^A_{m\gg1} =  -\frac{76423251279375 \sqrt{\pi } m^{13/2}}{34359738368\cdot 2^{\frac13}}
 \,.
}
While we have not been able to find $\hat d_m$ or $\hat {\hat d}_m$ in a closed form, we again find that they are sub-leading at large $m$. From the asymptotic behavior of ${\hat c}_m$ and $\hat{\hat c}_m$ we see that $M^{R|R^4}(s,t)$ and $M^{R^4|R^4}(s,t)$ are divergent as written. These divergences can be removed by adding the relevant contact term ambiguities in \eqref{M2222} at each order in $c$ with the appropriate infinite coefficient. These contact terms only affect low spins that we will not consider, so we do not write them here explicitly.

\subsection{Comparison to 11d}
\label{flatLimit}

We will now compare the 1-loop Mellin amplitudes to the corresponding M-theory amplitudes in 11d using the flat-space limit formula \eqref{flat} for $p=2$. To apply this formula to the 1-loop amplitudes, we should look at the regime where $m,n,s,t,u$ all scale equally large, in which case we can replace the sums over $m,n$ by integrals. For instance, for $M^{R|R}$ in each theory we have
\es{largeRR}{
\lim_{s,t\to\infty}M^{R|R}(s,t)=\frac{1}{\mathfrak{o}}\left[\int_{0}^\infty dmdn\frac{c^A_{m,n\gg1}}{(s-2m)(t-2m)}+\int_{0}^\infty dm\frac{c^A_{m\gg1}}{s-2m}(t^{-1}+u^{-1})+\text{crossed}\right]\,,
}
where $c^A_{m,n\gg1}$ and $c^A_{m\gg1}$ were given in \eqref{largecmn} and the only difference between $A_{N-1}$ and $D_N$ is the usual orbifold factor $\mathfrak{o}$, since $d_{mn}$ and $d_m$ in \eqref{cmn} are subleading at large $m,n$. We can perform these integrals to get
\es{intmn}{
&\int_{0}^\infty dmdn\frac{c^A_{m,n\gg1}}{(s-2m)(t-2m)}=\frac{36465 \pi ^{3/2}}{131072 \sqrt{2} (-s-t)^5 \sqrt{s t}} \Big(8 s^4 \sqrt{-t}-88 s^3 (-t)^{3/2}\\
   &\quad-105 s^2 t^2
   \sqrt{-s-t} \log \left({\left(\sqrt{-s-t}+\sqrt{-s}\right)
   \left(\sqrt{-s-t}+\sqrt{-t}\right)}(s t)^{-\frac12}\right)+41 s^2 (-t)^{5/2}+8
   \sqrt{-s} t^4\\
   &\quad-88 (-s)^{3/2} t^3+41 (-s)^{5/2} t^2+45 (-s)^{7/2} t+45 s
   (-t)^{7/2}-6 (-s)^{9/2}-6 (-t)^{9/2}\Big)\,,
}
and 
\es{intm}{
\int_{0}^\infty dm\frac{c^A_{m\gg1}}{s-2m}(t^{-1}+u^{-1})=-\frac{328185 \pi ^{3/2}}{32768 \sqrt{2} \sqrt{-s} }(t^{-1}+u^{-1})\,.
}
We can similarly take the large $s,t$ limit of $M^{R|R^4}$ and $M^{R^4|R^4}$ and use the expressions for $\hat c^A_{m\gg1}$ and $\hat {\hat c}^A_{m\gg1}$ in \eqref{largemRR4} to get\footnote{These integrals are divergent just as the sums were and must be regularized. As we discussed before for the sums, the regularization takes the form of the contact term ambiguities in \eqref{M2222} that are sub-leading at large $s,t$ and so can be ignored in the flat space limit.}
\es{largeRR4}{
\lim_{s,t\to\infty}M^{R|R^4}(s,t)&=\frac{1}{\mathfrak{o}^{\frac53}}\int_{0}^\infty dm\frac{\hat c^A_{m\gg1}}{(s-2m)tu}+\text{crossed}\\
&=-\frac{1}{\mathfrak{o}^{\frac53}}\frac{1673196525  \pi ^{3/2} (-s)^{7/2}}{134217728 \cdot 2^{\frac16} t u}+\text{crossed}\,,\\
}
and
\es{largeR4R4}{
 \lim_{s,t\to\infty}M^{R^4|R^4}(s,t)&=\frac{1}{\mathfrak{o}^{\frac73}}\int_{0}^\infty dm\frac{\hat {\hat c}^A_{m\gg1}}{(s-2m)tu}+\text{crossed}\\
 &=-\frac{1}{\mathfrak{o}^{\frac73}}\frac{76423251279375  \pi ^{3/2} (-s)^{13/2}}{4398046511104\cdot 2^{5/6}  tu}+\text{crossed}\,,
}
where again the only difference between the $A_{N-1}$ and $D_N$ theories is the powers of $\mathfrak{o}$ that come from \eqref{RR4cm} once we neglect $\hat d_m$ and $\hat{\hat d}_m$ since they are sub-leading at large $m$. 

We will now compare to the M-theory amplitude $\cA$ in 11d flat space, and we will normalize all our amplitude by tree supergravity $\cA_R$ as in the flat space limit formula \eqref{flat}. The 1-loop supergravity term $\ell_{11}^{18}\cA^{R|R}$ can be written in terms of a regularized scalar box integral $I_\text{reg}(s,t)$ as \cite{Russo:1997mk,Green:1997as}
\es{ARR}{
\frac{\cA_{R|R}}{\cA_R}=16\pi^5 stu(I_\text{reg}(s,t)+\text{crossed})\,,
}
where the prefactor comes from the gravitational constant $\kappa_{11}^2=16\pi^5 \ell_{11}^9$ in 11d. The unregularized 11d scalar box integral $I(s,t)$ is as usual  
   \es{dbox}{
   I(s,t)=&\int \frac{d^{11}q}{i(2\pi)^{11}}\frac{1}{q^2}\frac{1}{(q+p_1)^2}\frac{1}{(q+p_1+p_2)^2}\frac{1}{(q-p_4)^2}\,,
   }
   where the Mandelstam variables are related to the external momenta $p_i$ as $s=(p_1+p_2)^2$ and $t=(p_2+p_3)^2$. We regularized the box integral using a cutoff and performed the integrals to get
   \es{IR}{
    I_\text{reg}(s,t)=& \frac{1}{5898240\pi^4 (-s-t)^{7/2}} \big(  -15 (s t)^{5/2} \log
   \left({\left(\sqrt{-s-t}+\sqrt{-s}\right)
   \left(\sqrt{-s-t}+\sqrt{-t}\right)}{(s t)^{-\frac12}}\right) \\
   &+  \sqrt{-s-t} (23 s^2 (-t)^{5/2}+23 (-s)^{5/2} t^2-11
   (-s)^{7/2} t-11 s (-t)^{7/2} +3 (-s)^{9/2}+3 (-t)^{9/2})
  \big)\,,
   }
   where the regularization gives a cutoff dependent term that M-theory fixes in terms of $\ell_{11}^{15}\cA_{R^4}$ in \eqref{SGtoR4} as shown in \cite{Russo:1997mk,Green:1997as}. We can then compare $\frac{\cA_{R|R}}{\cA_R}$ in \eqref{ARR} to $\lim_{s,t\to\infty}M^{R|R}$ using the flat space limit formula \eqref{flat} (for $p=2$ and $a=5.5$) and the relation between $\ell_{11}$ and $c$ in \eqref{cPlanck}, and find a precise match. Note that the $\mathfrak{o}$ dependence in \eqref{cPlanck} for $\ell_{11}^9$ exactly matches that in \eqref{ARR}, so we have a check of $M^{R|R}$ for both $A_{N-1}$ and $D_N$.
   
 For the higher derivative 1-loop amplitudes $\ell_{11}^{24}\cA_{R|R^4}$ and $\ell_{11}^{30}\cA_{R^4|R^4}$, we can use the unitarity cut method to compute them in terms of their tree vertices, as was done for string theory in 10d flat space in \cite{Green:2008uj}. As we show in Appendix \ref{11dcut},  the $s$-channel formula for the discontinuity of a 1-loop amplitude $\mathcal{A}_{A|B}$ with vertices $A,B$ for 11d is
   \es{cut2}{
  \frac{\text{Disc}_s \mathcal{A}_{A|B}}{\mathcal{A}_{R}}=&-(2-\delta_{A,B})\frac{i  s^{15/2} tu}{393216}\int_0^\pi d\theta\int_0^{2\pi} d\phi \frac{\sin^8\theta |\sin^7\phi| }{s t't''(s+t')(s+t'')}\frac{\cA_A(s,t')}{\cA_R(s,t')} \frac{\cA_B(s,t'')}{\cA_R(s,t'')}\,,
   }
where $t',t''$ are defined in terms of $s,t,\theta,\phi$ in the Appendix. For $R|R$, this matches the discontinuity of \eqref{ARR}, as we show in the Appendix. For $\mathcal{A}_{R|R^4}$ and $\mathcal{A}_{R^4|R^4}$, we can do the integrals, use the fact that $\text{Disc}(\sqrt{s})=2\sqrt{s}$, and add the crossing symmetric terms to get
  \es{R4SGfin}{
  \frac{ \mathcal{A}_{R|R^4}}{\mathcal{A}_{R}}=\frac{ \pi  \ell_{11}^{15} (-s)^{11/2} t (s+t)}{66060288}+\text{crossed}\,,
   }
   and 
      \es{R4R4fin}{
 \frac{  \mathcal{A}_{R^4|R^4}}{\mathcal{A}_{R}}=\frac{ \pi  \ell_{11}^{21} (-s)^{17/2} t (s+t)}{231928233984}+\text{crossed}\,.
   }
We can then compare $\frac{\cA_{R|R^4}}{\cA_R}$ in \eqref{R4SGfin} and $\frac{\cA_{R^4|R^4}}{\cA_R}$ in \eqref{R4R4fin} to $\lim_{s,t\to\infty}M^{R|R^4}$ and $\lim_{s,t\to\infty}M^{R^4|R^4}$, respectively, using the flat space limit formula \eqref{flat} (for $p=2$ and $a=8.5,11.5$, respectively) and the relation between $\ell_{11}$ and $c$ in \eqref{cPlanck}, and find a precise match in each case.

\subsection{Extracting CFT data}
\label{CFTData}

Lastly, we can extract all low-lying CFT data from the $R|R$, i.e. $c^{-2}$, and $R|R^4$, i.e. $c^{-\frac83}$, correlators using two methods. Firstly, we derive an inversion integral formula for each DD in position space, which allows us to efficiently extract all CFT data above a certain spin, as expected from the Lorentzian inversion formula \cite{Caron-Huot:2017vep}. Secondly, we expand each entire correlator as written in Mellin space in conformal blocks to extract all CFT data for all spins up to the contact term ambiguities that appear in \eqref{M2222}. We find that both methods agree in their respective regimes of applicability. We do not extract CFT data from the $R^4|R^4$, i.e. $c^{-\frac{10}{3}}$, correlator, since we anyway do not know the 2-loop term $R|R|R$ that would contribute at the earlier order $c^{-3}$. Nevertheless, from the formula we present it would be simple to extract the $R^4|R^4$ data as well if desired.

First we describe how to extract CFT data from the DD. For this purpose we are interested in the behaviour of the small $z$ slices in \eqref{slices2} as expanded around $\bar z=1$. In this limit all slices behave as
\begin{equation}
h^{(n)}(\bar z)  = h_\text{no-log}^{(n)} (\bar z)+   (1-\bar z) \log(1-\bar z) h_\text{log}^{(n)}(\bar z)+ \cdots\,.
\end{equation}
The coefficients $h_\text{no-log}^{(n)}$ and $h_\text{log}^{(n)}$ can be found in a closed form in all cases, but their expression is not very illuminating. Upon $1\leftrightarrow3$ crossing symmetry \eqref{crossing} these slices give the piece of the correlator proportional to $\log^2V$ in a small $V$ expansion. More precisely, for each contribution the $t$-channel expression for any 1-loop term $A|B$ is
\es{sing}{
{\cal H}^{A|B}(U,V)&= \frac{U^4}{V^4} \log^2 V \left( (1-\bar z)^5 h_{A|B}^{(5)}(1-z) + \cdots  \right) + \cdots\,,\\
&= z^4 \log^2 (1-\bar z) \left( h^{A|B}_\text{no-log}(\bar z)+h^{A|B}_\text{log}(\bar z)z\log z  \right) + \cdots\,,\\
}
where in the second line we expanded for small $z$ and defined the leading order resummed slices
\es{defh}{
h^{A|B}_\text{no-log}(\bar z)\equiv \bar z^4\sum_{n=5}^\infty(1-\bar z)^n h_{\text{no-log}}^{A|B,(n)}(1)\,,\qquad h^{A|B}_\text{log}(\bar z)\equiv \bar z^4\sum_{n=5}^\infty(1-\bar z)^n h_\text{log}^{A|B,(n)}(1)\,.
}
We focus on the $\log^2 (1-\bar z)$ term because it is the only contribution to the DD, which takes the form
\es{DD}{
{\rm dDisc}\,[ f(\bar z) \log^2(1-\bar z) ] = 4\pi^2  f(\bar z)\,,
}
for arbitrary $f(\bar z)$ analytic at $\bar z=1$. We will compare this DD to the $s$-channel block expansion of ${\cal H}^{A|B}(U,V)$ in the small $z$, i.e. $U$, expansion. The main contribution arises from the tower of operators ${\cal B}[0,2]_\ell$, with odd spin, and the scalar operator ${\cal D}[04]$. The DD can only be used to extract CFT data with $\ell>0$ for $R|R$, so we will focus on $\lambda^2_{\cB[02]_\ell}$  at these orders. From \eqref{RR} we see that their contribution to the reduced correlator to leading order in $z$ at any order beyond tree level is
\es{smallU}{
\left. {\cal H}(U,V) \right|_{{\cal B}[0,2]} = (z\bar z)^4 \sum_{\ell=1,3,\cdots} \lambda^2_{\cB[02]_\ell}  C_{\cB[02]_\ell }g_{\ell+11,\ell+1}^{0,-2}(\bar z) + { O}(z^5)\,,
}
where the lightcone block $g_{\Delta,\ell}^{0,-2}(\bar z)$ is the leading small $z,\bar z$ term in the conformal block $G^{0,2}_{\Delta,\ell}(U,V)$ as described in Appendix \ref{block6dApp}. We can then extract $ \lambda^2_{\cB[02]_\ell}  $ by requiring that the sum in \eqref{smallU} matches the singularity in \eqref{sing}. This can be done by applying the Lorentzian inversion formula \cite{Caron-Huot:2017vep} to the DD in a small $z$ expansion. As we show in Appendix \ref{largeSpinApp}, the inversion integral in our case can also be derived from large spin perturbation theory \cite{Alday:2016njk,Alday:2015eya} and takes the form
\es{inversion}{
\lambda^2_{\cB[02]_\ell}  =  -\frac{2^{-4 \ell-19}\pi (\ell+1) (\ell+4) \Gamma (\ell+5) \Gamma (\ell+7)}{  (\ell+2) \Gamma \left(\ell+\frac{11}{2}\right) \Gamma \left(\ell+\frac{13}{2}\right)} \int_0^1 d\bar z  \bar z^{4} g_{\ell+11,\ell+1}^{-2,0}(\bar z) h_\text{no-log}(\bar z)\,,
}
where the normalization was fixed by demanding that the GFFT term, i.e. $c^0$, term in \eqref{treedata2} matches the appropriate GFFT singularity, as explained in the Appendix.  For $h^{R|R}_\text{no-log}$ this integral converges for all $\ell\geq1$, while for $h^{R|R^4}_\text{no-log}$ it converges for $\ell\geq5$, which as expected are precisely the spins that are not affected by the counterterms in Table \ref{resultList}.

We will also be interested in extracting the anomalous dimensions of the double-trace ${\cal L}[00]_{t+\ell,\ell}$ of leading twist $t=8$. From \eqref{RR} we see that their contribution to $\cH^{R|R}$ to leading order in $z$ can be found by looking at the $\log z$, i.e. $\log U$, coefficient
\es{smallUAnom}{
\left. {\cal H}^{R|R}(U,V) \right|_{\log z} &=  \frac{(z\bar z)^5}{2}\sum_{\ell=0,2,\cdots}\Big[(\lambda^{(0)}_{8,\ell})^2\gamma^{{R}|{R}}_{8,\ell}+(\lambda^{R})^2_{t,\ell}\gamma^{R}_{t,\ell}\\
&\qquad\qquad+(\lambda^{(0)}_{t,\ell})^2(\gamma^R_{t,\ell})^2\partial_t^\text{no-log}\Big]C_{t+\ell,\ell} g_{t+\ell+4,\ell}^{0,-2}(\bar z)\big\vert_{t=8} + { O}(z^6)\,,
}
while from \eqref{RR4} we have the similar formula
\es{smallUAnomR4}{
 &{\cal H}^{R|R^4}(U,V) |_{\log z} =  \frac{(z\bar z)^5}{2}\Big[\sum_{\ell=0,2,\cdots} (\lambda^{(0)}_{8,\ell})^2\gamma^{{R}|{R^4}}_{8,\ell}C_{8,\ell} g_{12+\ell,\ell}^{0,-2}(\bar z)\\
&+[(\lambda^{R})^2_{8,0}\gamma^{R^4}_{8,0}+(\lambda^{R^4})^2_{8,0}\gamma^{R}_{8,0}+2(\lambda^{(0)}_{8,0})^2(\gamma^{R^4}_{8,0})^2\partial_t^\text{no-log}]C_{t,0} g_{t+4,0}^{0,-2}(\bar z)\big\vert_{t=8} \Big]+ { O}(z^6)\,.
}
There are two subtleties when extracting the 1-loop anomalous dimensions from these formula, which both only affect $R|R$. Firstly, the inversion formula leads to CFT data written in terms of the conformal spin $J_\text{conf}^2 ={(2+\frac{\Delta+\ell}{2}) (1+\frac{\Delta+\ell}{2})}$, as opposed to the spin. In these variables, the twist for $t=8$ long operators in \eqref{anomLowTwist} with $\ell>0$ takes the form 
\begin{equation}
\label{reciprocity}
\text{twist} = 8 - c^{-1} \frac{17280}{(J_\text{conf}^2-12)(J^2_\text{conf}-20)} + c^{-2} \hat \gamma^{R|R}_{8,\ell} + \cdots\,\,,
\end{equation}
where $J^2_\text{conf}$ has an expansion in $c$ due to its dependence on $\Delta_{8,\ell}$, so that $\hat \gamma^{R|R}_{8,\ell}$ is related to $ \gamma^{R|R}_{8,\ell}$ as
\es{hatNohat}{
\gamma^{R|R}_{8,\ell}=\hat\gamma^{R|R}_{8,\ell}+\gamma^\text{extra}_\ell\,,\qquad\gamma^\text{extra}_\ell\equiv -\frac{298598400 (2 \ell+11) \left(\ell^2+11 \ell+14\right)}{(\ell+1)^3 (\ell+2)^3 (\ell+9)^3 (\ell+10)^3}\,.
}
We can then use this equation to write \eqref{smallUAnom} in terms of $ \hat\gamma^{R|R}_{8,\ell}$, so that an extra $(\lambda^{(0)}_{8,\ell})^2\gamma^\text{extra}_{\ell}$ will appear. The second subtlety is that since we are only interested in extracting $\gamma^{R|R}_{8,\ell}$, we must subtract off the DD for the other terms that depend on GFFT and tree level data, as well as $(\lambda^{(0)}_{8,\ell})^2\gamma^\text{extra}_{\ell}$, which is\footnote{The full sum takes the form 
\begin{eqnarray*}
&h_\text{extra}(\bar z)\log^2(1-\bar z)+ \frac{p^{(4)}(\bar z)}{\bar z^{10}}\text{Li}_3(1-\bar z) + \frac{p^{(4)}(\bar z)}{\bar z^{10}}\text{Li}_3(\bar z) + \frac{p^{(4)}(\bar z)}{\bar z^{10}}\text{Li}_2(1-\bar z)\log(1-\bar z) + \frac{p^{(4)}(\bar z)}{\bar z^{10}}\text{Li}_2(1-\bar z) \log(\bar z)  \\
&+ \frac{p^{(3)}(\bar z)}{\bar z^{9}}\log(1-\bar z) \log^2 \bar z+ \frac{p^{(3)}(\bar z)}{\bar z^{9}}\log(1-\bar z) \log \bar z+\frac{p^{(4)}(\bar z)}{\bar z^{9}}\log \bar z+\frac{p^{(5)}(\bar z)}{\bar z^{10}}\log(1- \bar z)+\frac{p^{(4)}(\bar z)}{\bar z^{10}},
\end{eqnarray*}
where $p^{(n)}(\bar z)$ denotes a polynomial of degree $n$. 
}
\es{DDextra}{
h_\text{extra}(\bar z)&\equiv  \frac{\bar z^5}{2}\sum_{\ell=0,2,\cdots}\Big[[(\lambda^{(0)}_{8,\ell})^2\gamma^\text{extra}_{\ell}+(\lambda^{R})^2_{8,0}\gamma^{R}_{8,0}\\
&\qquad\qquad\quad+(\lambda^{(0)}_{8,0})^2(\gamma^{R}_{8,0})^2\partial_t^\text{no-log}]C_{t,0} g_{t+4,0}^{0,-2}(\bar z)\big\vert_{t=8} \Big]_{\log^2(1-\bar z)}\\
&=  \frac{3456 (1-\bar z) \left(397 \bar z^3-2910 \bar z^2+5730 \bar z-3305\right)}{\bar z^5}\,.
}
We can now write an inversion integral for $\hat \gamma^{R|R}_{8,\ell}$, similar to that of $\lambda^2_{\cB{02}_\ell}$ described above, which takes the form
\es{hat}{
\hat \gamma^{R|R}_{8,\ell} = -\frac{\sqrt{\pi}45\ 2^{-7-2\ell} \Gamma (\ell+5)}{ (\ell+1) (\ell+2) (\ell+9) (\ell+10) \Gamma \left(\ell+\frac{13}{2}\right)}  \int_0^1 d\bar z \bar z^{4} g_{\ell+11,\ell+1}^{-2,0}(\bar z) (h_{\log }(\bar z)-h_\text{extra}(\bar z))\,,
}
where the normalization was again fixed by demanding that the GFFT twist, i.e. $8$, matches the appropriate tree singularity, as explained in the Appendix. We can then get $\gamma^{R|R}_{8,\ell}$ from $\hat\gamma^{R|R}_{8,\ell}$ using \eqref{hatNohat}. For $R|R^4$, we do not have these subtleties, since tree $R^4$ data is only nonzero for zero spin, so the extra terms in \eqref{smallUAnomR4} do not contribute to the DD, and the expansion in conformal spin is the same as spin for $\ell>0$. We then get a simpler inversion integral
\es{hatR4}{
 \gamma^{R|R^4}_{8,\ell} = -\frac{\sqrt{\pi}45\ 4^{-4-\ell} \Gamma (\ell+5)}{ (\ell+1) (\ell+2) (\ell+9) (\ell+10) \Gamma \left(\ell+\frac{13}{2}\right)}  \int_0^1 d\bar z \bar z^{4} g_{\ell+11,\ell+1}^{-2,0}(\bar z) h^{R|R^4}_{\log }(\bar z)\,,
}
which will only converge for $\ell\geq6$ as expected.

We can now expand the $t$-channel DD in \eqref{sing} in $z$, which is equivalent to expanding the $s$-channel slices $h^{(n)}(\bar z)$ around $\bar z=1$ and keeping the contributions relevant to computing the CFT-data for the leading twist operators discussed above, which we describe explicitly for each theory and 1-loop amplitude in Appendix \ref{smallDD}. We can plug these expressions into the inversion integrals derived above to get the low-lying $A_{N-1}$ data:
\es{BA}{
A_{N-1}:\qquad( \lambda^{R|R}_{{\cal B}[02]_1})^2&= \frac{1314669460}{231}-\frac{6738525487050 \pi ^2}{11685817} \,, \\
 ( \lambda^{R|R}_{{\cal B}[02]_3})^2&= \frac{104148260864}{1925}-\frac{997945007151708 \pi ^2}{182047645}\,,\\
( \lambda^{R|R}_{{\cal B}[02]_5})^2&=\frac{79969221983613084}{554498945}-\frac{3089164599444721545 \pi ^2}{211406651834}\,,\\
( \lambda^{R|R^4}_{{\cal B}[02]_5})^2&=\frac{54100697319516717819180\pi ^2}{79816797121\cdot 2^{\frac23}}-\frac{2162942188071237504}{323323\cdot 2^{\frac23}}\,,\\
( \lambda^{R|R^4}_{{\cal B}[02]_7})^2&=\frac{7005977256176634386899248000  \pi ^2}{3236491306459429\cdot 2^{\frac23}}-\frac{129989503159386400000}{6084351\cdot 2^{\frac23}}\,,\\
\gamma_{8,2}^{R|R}&=\frac{296924982566840}{27951}-\frac{33159966691580505 \pi ^2}{30808063}\,, \\
\gamma_{8,4}^{R|R} &= \frac{55393629950045341536}{290124835}-\frac{2654868179611476180 \pi ^2}{137235917}\,,\\
\gamma_{8,6}^{R|R}&=\frac{480348794643838967321}{247206960}-\frac{2460777825133887825423 \pi ^2}{12499025056}\,,\\
\gamma_{8,6}^{R|R^4}&=\frac{7602699448011614165949980250 \pi ^2}{435656388001\cdot 2^{\frac23}}-\frac{13262158264899710400}{77\cdot 2^{\frac23}} \,, \\
\gamma_{8,8}^{R|R^4}&=\frac{5601117586403483431983154368000 \pi ^2}{20475850236047 \cdot 2^{\frac23}}-\frac{4246794181892701017600}{1573 \cdot 2^{\frac23}}\,,
}
where for $R|R$ we can only compute spin $\ell>0$, and for $R|R^4$ we could only compute $\ell>4$. For $D_N$, we could not compute the integrals in closed form, but they can be evaluated numerically to any precision to get:
\es{BD}{
D_{N}:\qquad ( \lambda^{R|R}_{{\cal B}[02]_1})^2&\approx  -7.62943\,, \qquad\quad
 ( \lambda^{R|R}_{{\cal B}[02]_3})^2\approx-0.15983\,,\qquad
 ( \lambda^{R|R}_{{\cal B}[02]_5})^2\approx -0.006606\,,\\
  ( \lambda^{R|R^4}_{{\cal B}[02]_5})^2&\approx-1.1105914\,,\qquad
( \lambda^{R|R^4}_{{\cal B}[02]_7})^2 \approx-0.0209759\,,\\
 \gamma_{8,2}^{R|R}&\approx-644.2556829 \,,\qquad\;\;\;
 \gamma_{8,4}^{R|R}\approx-18.9188879\,,\qquad\;
 \gamma_{8,6}^{R|R}\approx-2.585552637\,,\\
  \gamma_{8,6}^{R|R^4}&\approx -1119.953\,,\qquad\qquad
\gamma_{8,8}^{R|R^4}\approx -86.69989\,.
}

Finally, we can also extract CFT from the entire correlator as written in Mellin space in terms of the contact term ambiguities described before. We extract this data by simply expanding in blocks. Recall that poles in the Mellin amplitude correspond to twists of operators, where we should consider the twists of multiplets in the reduced correlator as summarized in Table \ref{moreMult}. The lowest twists of operator that appear at 1-loop are the $\cB[02]_\ell$ and $\cD[04]$, which both have effective twist 6. In the $s$-channel, these correspond to $(s-6)^{-1}$ terms in the Mellin amplitudes of the previous subsection.\footnote{In $\mathcal{N}=4$ SYM, the analogous 1-loop amplitudes in \cite{Alday:2018kkw} did not contain such terms, since in that case the only operators that appear at 1-loop are the unprotected double trace operators.} The lowest twist double trace anomalous dimensions similarly correspond to $(s-8)^{-2}$ terms. 

We can extract CFT data in each case by taking the relevant $s$-pole, doing the $t$-integral, and then projecting against a block of the corresponding spin using the projectors introduced in \cite{Heemskerk:2009pn}. This method of extracting CFT data was applied to maximal SUSY holographic theories in \cite{Chester:2018lbz}, and we give the details for our very similar case in Appendix \ref{extraction}. We used this method to reproduce all the CFT computed for both $A_{N-1}$ and $D_N$ using the DD method. We furthermore computed $\gamma_{8,0}^{R|R}$ and $(\lambda_{\cD[04]}^{R|R})^2$, which could not be computed from the DD. To these contributions we should add that of the contact term $M^4$ in Table \ref{moreMult}, so that the total answer at each order is:
\es{spin0A}{
A_{N-1}:\qquad \gamma_{8,0}^{R|R}&=16897.3-\frac{360}{11}B_4^{A}\,,\\
(\lambda_{\cD[04]}^{R|R})^2&=\frac{21886734}{245}-\frac{13669988758155 \pi ^2}{1513521152}- \frac{3}{7}B_{4}^{A}\,,\\
D_{N}:\qquad \gamma_{8,0}^{R|R}&=15588.3-\frac{360}{11}B_4^{D}\,,\\
(\lambda_{\cD[04]}^{R|R})^2&=130.383- \frac{3}{7}B_4^{D}\,,\\
}
where $B_4^{A}$ and $B_4^{D}$ are the coefficients of the contact term ambiguity in \eqref{M2222} for $A_{N-1}$ and $D_N$, respectively, and the data shown numerically can be computed to any desired precision. One could similarly extract the CFT data from $R|R^4$ for $\ell\leq4$ in terms of the four contact term ambiguities at order $c^{-\frac83}$ in \eqref{M2222}, by first regularizing $M^{R|R^4}$ as in \cite{Chester:2019pvm} and then applying the projection method.

\section{Conclusion}
\label{conc}

In this paper we considered the large $c$ expansion of the stress tensor multiplet correlator in both the $A_{N-1}$ and $D_N$ $(2,0)$ theories and computed the corrections coming from 1-loop terms with either supergravity $R$ or the first higher derivative $R^4$ vertices. For the $A_{N-1}$ theory, we wrote the complete expression for these terms in Mellin space, up to contact term ambiguities, extracted the low lying CFT data from them, and checked that they precisely matched the corresponding terms in the 11d M-theory S-matrix in the flat space limit. For the $D_N$ theory, we could not find the complete closed form expression for the correlator, but nevertheless we computed enough terms to similarly check the flat space limit and extract CFT data, just as we did for $A_{N-1}$. This computation is not only the first example of a 1-loop correction for a non-Lagrangian theory, but it is also the first precision check of AdS/CFT for 1-loop higher derivative terms. 

Our results imply that all nonzero spin CFT data is bigger for the $D_N$ theory at large enough $c$ where the $O(c^{-2})$ approximation is accurate, while for zero spin CFT data the $D_N$ theory is bigger where the $O(c^{-\frac53})$ approximation is accurate, since the $R^4$ contact term only contributes to zero spin data. As discussed in the Introduction, the numerical bootstrap study in \cite{Beem:2015aoa} showed that at large $c$ the bootstrap bounds were approximately saturated by the $c^{-1}$ correction, which is the same for both $A_{N-1}$ and $D_N$. Our results show that if either theory truly saturates the bounds, then it must be the $D_N$ theory. Since there is a unique solution to the crossing equations at the boundary of the allowed region, this conjecture would mean that the numerical bootstrap provides a non-perturbative solution of the $D_N$ theory for all $N$, and all CFT data in the correlator could be read off from this solution using for instance the extremal functional method \cite{Poland:2010wg,ElShowk:2012hu,El-Showk:2014dwa}. Unfortunately, it is difficult to verify this conjecture with the $O(c^{-2})$ approximation computed here, for the following reasons. The lowest interacting $D_N$ theory is $D_2$, but this is dual to $A_1\times A_1$, and so is a product theory that cannot saturate bootstrap bounds \cite{Chester:2014mea}.\footnote{For instance, the scaling dimension upper bound for $D_2$ would have to be at least as restrictive as $A_1$. We thank Ofer Aharony for discussions about this.} The next lowest $D_N$ theory is $D_3$. but this is dual to $A_3$, so the lowest $D_N$ theory that we could use to compare numerics to analytics is $D_4$ with $c(D_4)=676$. For this very large value of $c$, we would need to compute CFT data to approximately 6 digits of accuracy to detect the tiny $c^{-2}$ correction that distinguishes $A_{N-1}$ and $D_N$, which is especially difficult considering that the 6d bootstrap is already much less numerically stable than lower dimensions bootstrap studies.

Another aspect of our 1-loop computation that we would like to emphasize is that it was not necessary to completely unmix the degenerate double-trace operators. Instead, we only needed to compute the average of the anomalous dimension squared starting from the average of the anomalous dimension, which requires a partial unmixing. This was in fact the same in the previously considered $\cN=4$ SYM case. One could further try to completely unmix the double-trace operators, as was done for $SU(N)$ $\cN=4$ SYM in \cite{Aprile:2017xsp}, and the result took a remarkably simple form of rational numbers. This simplicity was explained in \cite{Caron-Huot:2018kta} as arising from a hidden 10d conformal symmetry at this order in the large $c$ expansion, which may arise from the fact that $AdS_5\times S^5$ is conformally flat. In our 6d case, neither $AdS_7\times S^4$ nor $AdS_7\times S^4/\mathbb{Z}_2$ for the $A_{N-1}$ and $D_N$ theories, respectively, is conformally flat. Indeed, one can try to perform the complete unmixing of some of the low twist double trace operators for $A_{N-1}$ and $D_N$ using the tree $\langle22pp\rangle$ and GFFT $\langle pppp\rangle$ data computed in this work, and one will quickly see that the unmixed operators have ugly anomalous dimensions involving complicated roots that one generically expects from diagonalizing arbitrary matrices. This suggests that while this hidden conformal symmetry may be useful for performing the complete unmixing, it is not necessary for computing higher loop correlators.

Our results open the door for a more extensive study of 6d $(2,0)$ CFTs at 1-loop for other correlators $\langle ppqq\rangle$ with $q\geq p$. As discussed, the most difficult part of such 1-loop computations is the GFFT average OPE coefficients, which we presented here for general $\langle pppp\rangle$. To compute other correlators at 1-loop, as has been done for $\cN=4$ SYM in \cite{Alday:2019nin,Aprile:2017qoy,Aprile:2019rep}, one would merely need to compute the average anomalous dimensions, which is easier since one only needs to consider the superblock expansion for $\langle qqrr\rangle$ for fixed $q$ and all $r$. One difficulty for correlators with $q>2$ is that a superblock expansion for the reduced correlator is only known for $q=2$, which should make the calculation more challenging, but not impossible. We hope to report back on results for $\langle3333\rangle$ soon.

It would also be nice to compute 1-loop holographic correlators in other dimensions. In 4d $\cN=4$ SYM, the 1-loop terms for the $SU(N)$ theory, which is dual to $AdS_5\times S^5$, have been computed, but not for the $SO(2N)$ theory, which is dual to the orbifold $AdS_5\times S^5/\mathbb{Z}_2$, and so is analogous to the $D_N$ calculation in this work. Also, 1-loop terms with $R^4$ vertices were considered in \cite{Alday:2018pdi,Drummond:2019hel}, but they have been matched to the flat space limit only up to an overall factor, unlike in this work, where the match is precise. In 3d, no 1-loop holographic correlators have been computed, although genus-one contact terms were derived using localization in \cite{Binder:2019mpb,Binder:2018yvd,Chester:2018aca}. In 2d, $\langle22pp\rangle$ at tree level was recently computed for a certain half-maximal susy CFT with a holographic dual \cite{Giusto:2018ovt,Giusto:2019pxc,Rastelli:2019gtj}, but the 1-loop term has not yet been computed.

\newpage
\section*{Acknowledgments} 

We thank Ofer Aharony, Silvu Pufu, Michael Green, Chris Beem, Balt van Rees, David Simmons-Duffin, Thomas Dumitrescu, David Poland, Walter Landry, and Xinan Zhou for useful conversations, and Ofer Aharony for reading through the manuscript. The work of LFA is supported by the European Research Council (ERC) under the European Union's Horizon
2020 research and innovation programme (grant agreement No 787185). SMC is supported by the Zuckerman STEM Leadership Fellowship. HR acknowledges the support from the PBC postdoctoral fellowship program as well as the Israel Science Foundation center for excellence grant (grant number 1989/14) and by the Minerva foundation with funding from the Federal German Ministry for Education and Research. The authors would like to acknowledge the use of the University of Oxford Advanced Research Computing (ARC) facility in carrying out this work.(http://dx.doi.org/10.5281/zenodo.22558)
\appendix

\section{6d Conformal blocks}
\label{block6dApp}
In this appendix we give explicit formulae for the 6d conformal blocks and their expansion in various useful variables. The 6d conformal block that appears in a four-point function of operators of dimension $\Delta_i$, $(i=1,2,3,4)$ has been obtained in closed form in \cite{Dolan:2003hv,Dolan:2011dv}:
\es{blockApp1}{
&G_{\Delta,\ell}^{\Delta_{12},\Delta_{34}}(z,\bar{z})=\cF_{0,0}-\frac{(\ell+3)}{\ell+1}\cF_{-1,1}+\frac{2 (\Delta -4) \Delta_{12} \Delta_{34} (\ell+3)}{(\Delta +\ell) (\Delta +\ell-2) (\Delta -\ell-4) (\Delta -\ell-6)}\cF_{0,1}\\
&+\frac{(\Delta -4) (\ell+3) (\Delta -\Delta_{12}-\ell-4) (\Delta +\Delta_{12}-\ell-4) (\Delta +\Delta_{34}-\ell-4) (\Delta -\Delta_{34}-\ell-4) }{16 (\Delta -2) (\ell+1)(\Delta -\ell-5) (\Delta -\ell-4)^2 (\Delta -\ell-3) }\cF_{0,2}\\
&-\frac{(\Delta -4) (\Delta -\Delta_{12}+\ell) (\Delta +\Delta_{12}+\ell) (\Delta +\Delta_{34}+\ell) (\Delta -\Delta_{34}+\ell)}{16(\Delta -2) (\Delta +\ell-1) (\Delta +\ell)^2 (\Delta +\ell+1)}\cF_{1,1}\,,\\
}
where we define
\es{blockApp2}{
&\mathcal{F}_{nm}(z,\bar z)\equiv \frac{(z \bar z)^{\frac{\Delta-\ell}{2}}}{(z-\bar z)^3}\left({(-z)}^\ell z^{n+3}\bar z^m{}_2F_1\left(\frac{\Delta+\ell-\Delta_{12}}{2}+n,\frac{\Delta+\ell+\Delta_{34}}{2}+n,\Delta+\ell+2n;z\right)\right.\\
&\left. {}_2F_1\left(\frac{\Delta-\ell-\Delta_{12}}{2}-3+m,\frac{\Delta-\ell+\Delta_{34}}{2}-3+m,\Delta-\ell-6+2m;\bar z\right)-(z\leftrightarrow \bar z)\right)\,.
}
Our normalization for the blocks differs from those in \cite{Beem:2015aoa} by a factor of $2^\ell$. Also, here we have corrected a typo in that paper. 

We also find it convenient to expand these blocks in terms of Jack polynomials. This expansion takes the form \cite{Dolan:2003hv,Dolan:2011dv}:\es{blockApp3}{
&G_{\Delta,\ell}^{\Delta_{12},\Delta_{34}}(z,\bar{z})=(-1)^\ell\sum_{m,n=0}^\infty r_{m,n} ~P_{\frac12(\Delta+\ell)+m,\frac12(\Delta-\ell)+n}(z,\bar{z})\,,\\
}
where the Jack polynomials in 6d are defined as
\es{Jack}{
&P_{a,b}(z,\bar z)=\frac{6}{(z-\bar z)^3(2+a-b)}\left(\frac{z^{b+1}\bar z^{2+a}-z^{2+a}\bar z^{1+b}}{1+a-b}+\frac{z^{3+a}\bar z^b-z^b\bar z^{a+3}}{a-b+3}\right)\,,
}
and the coefficients $r_{m,n}$ are
\es{rmn}{
&r_{mn}=\frac{(\ell+m-n+2) (\Delta  (\ell+1) (\ell+m+3)+2 n (\ell-m+3)-2 (\ell+1) (\ell+m+3)-\Delta  (\ell+3) n)}{6m! n! (\Delta -2) (\ell+1) (\ell+\Delta )_m (-\ell+\Delta -4)_n}  \\
& \times \left(\frac{1}{2} (\ell+\Delta -\Delta_{12})\right)_m \left(\frac{1}{2} (\ell+\Delta +\Delta_{34})\right)_m \left(\frac{1}{2} (\Delta-\ell -\Delta_{12}-4)\right)_n \left(\frac{1}{2} (\Delta-\ell +\Delta_{34}-4)\right)_n\,.
}

There is a yet another convenient expansion of conformal blocks that is organized by twist. It is called the lightcone or collinear conformal block expansion and takes the form
\begin{align}
G^{\Delta_{12},\Delta_{34}}_{\Delta,\ell}(U,V)=\sum_{k=0}^\infty U^{\frac{\Delta-\ell}{2}+k}g_{\Delta,\ell}^{[k],\Delta_{12},\Delta_{34}}(V)\,,
\end{align}
where $U=z \bar{z}$ and $V=(1-z)(1-\bar{z})$. The lightcone blocks $g_{\Delta,\ell}^{[k],\Delta_{12},\Delta_{34}}(V)$ are labeled by $k$ and are only functions of $V$. We will only use the $k=0$ lightcone blocks, which for simplicity we denote without the $[0]$ subscript, and which take the form
\es{lightblock}{
&g_{\Delta,\ell}^{\Delta_{12},\Delta_{34}}(V)=(V-1)^\ell \,{}_2F_1\left(\frac{\Delta+\ell-\Delta_{12}}{2},\frac{\Delta+\ell+\Delta_{34}}{2},\Delta+\ell,1-V\right)\,.\\
}

The last conformal block expansion we used is the radial expansion, which is naturally organized by dimension. This expansion is written in terms of the radial coordinates $r,\eta$ introduced in \cite{Hogervorst:2013sma}, which are related to $z,\bar z$ as
\begin{align}
z=\frac{4 r \left(\eta +\sqrt{\eta ^2-1} \left(r^2-1\right)+\eta  r^2+2 r\right)}{\left(r^2+2 \eta 
   r+1\right)^2}~,\qquad \bar{z}= \frac{4 r \left(\eta -\sqrt{\eta ^2-1} \left(r^2-1\right)+\eta  r^2+2 r\right)}{\left(r^2+2 \eta 
   r+1\right)^2}\,.
\end{align}
We can derive an efficient recursion relation in terms of $r$ by noting that the exact conformal block \eqref{blockApp1} can be written in terms of the leading lightcone blocks $g_{\Delta,\ell}^{\Delta_{12},\Delta_{34}}(V)$ if we set $V=1-z$ or $V=1-\bar z$. Efficient recursion formulae have been worked out for $g_{\Delta,\ell}^{\Delta_{12},\Delta_{34}}(V)$ in for instance Appendix E of \cite{Karateev:2019pvw}, which when plugged back into \eqref{blockApp1} are organized naturally as an expansion in $r$ for finite $\eta$.

   \section{Unitarity cut in 11d}
\label{11dcut}

In this Appendix, we describe how higher derivative 1-loop amplitudes $\cA_{A|B}$ in 11d with vertices $A,B$ can be computed from the tree amplitudes $\cA_A$ and $\cA_B$ using unitarity. We will follow a similar derivation done for string theory in 10d in \cite{Green:2008uj}, but use the conventions of \cite{Bern:1998ug}.\footnote{Except that our $\kappa^2_{11}$ is four times smaller than that of \cite{Bern:1998ug}.}

We begin by defining the Einstein-Hilbert term as
  \es{EHD}{
   -\frac{1}{2\kappa_{11}^2}\int d^{D}x\sqrt{-g}R\,,\qquad \kappa_{11}^2=16\pi^5\ell_{11}^{9}\,,
   }
   where the value of the gravitational constant $\kappa_{11}^2$ follows \cite{Russo:1997mk}. Then the tree level supergravity term in the 11d amplitude \eqref{A} is
   \es{anyD}{
   \mathcal{A}_{R}=&16\pi^5\frac{16 i \bold{R}^2}{stu}\,,\\
   }
   where the polarization factor is
   \es{R}{
   \bold{R}=(t_8)_{\mu_1\nu_1\mu_2\nu_2\mu_3\nu_3\mu_4\nu_4}p_1^{\mu_1}p_2^{\mu_2}p_3^{\mu_3}p_4^{\mu_4}\epsilon_1^{\nu_1}\epsilon_2^{\nu_2}\epsilon_3^{\nu_3}\epsilon_4^{\nu_4}\,,
   }
   with $t_8$ a tensor defined in (9.A.18) of Green, Schwartz, Witten (after dropping the 9d Levi-Civita tensor).

We can now compute 1-loop terms $\cA_{A|B}$ with vertices $A,B$ by taking the two particle unitarity cut, which expresses the discontinuity of 1-loop amplitudes with momenta $p_i$ and polarization $\zeta_i$ in terms of tree level amplitudes $\cA_A$ and $\cA_B$ as (we consider just the $s$-channel here):
   \es{cut}{
 &  \text{Disc}_s\mathcal{A}^{A|B}_{\zeta_1,\zeta_2,\zeta_3,\zeta_4}(p_i)=-\frac{(2-\delta_{A,B})}{32}\int \frac{dq^{11}}{(2\pi)^{11}}2\pi \delta^{(+)}(q^2)2\pi \delta^{(+)}((p_1+p_2-q)^2)\\
   &\qquad\times\sum_{\{\zeta_r,\zeta_s\}}\mathcal{A}^{A}_{\zeta_1,\zeta_2,\zeta_r,\zeta_s}(p_1,p_2,-q,q-p_1-p_2)\mathcal{A}^{B}_{\zeta_r,\zeta_s,\zeta_3,\zeta_4}(p_3,p_4,q,p_1+p_2-q)\,,
   }
   where $(2-\delta_{A,B})$ takes into account the fact that $\mathcal{A}^{A|B}=\mathcal{A}^{B|A}$ appears twice, $\sum_{\{\zeta_r,\zeta_s\}}$ denotes the sum over all two-particle massless supergravity states, and $\delta^{(+)}(p^2)=\delta^{(11)}(p^2)\theta(p^0)$ imposes the mass-shell condition on each intermediate massless state:
   \es{massshell}{
   q^2=0\,,\qquad (p_1+p_2-q)^2=0\,.
   }
    The product of polarizations in \eqref{cut} can then be simplified from the so-called self-replicating formula that holds for all maximal supergravity theories:
   \es{selfrep}{
 \sum_{\{\zeta_r,\zeta_s\}}  \bold{R}^4_{\zeta_1,\zeta_2,\zeta_r,\zeta_s}(p_1,p_2,-q,q-p_1-p_2) \bold{R}^4_{\zeta_r,\zeta_s,\zeta_3,\zeta_4}(p_3,p_4,q,p_1+p_2-q)=\frac{s^4}{16} \bold{R}^4_{\zeta_1,\zeta_2,\zeta_3,\zeta_4}(p_1,p_2,p_3,p_4) \,.
   }
   We can now evaluate the 11d phase space integral by going to the center of mass frame
   \es{frame}{
   p_1^\mu&=\frac{\sqrt{s}}{2}\begin{pmatrix} 1&1&\vec0_9\end{pmatrix}\,,\qquad  p_2^\mu=\frac{\sqrt{s}}{2}\begin{pmatrix} 1&-1&\vec0_9\end{pmatrix}\,,\\
      p_3^\mu&=\frac{\sqrt{s}}{2}\begin{pmatrix} -1&\cos\rho&\sin\rho&\vec0_8\end{pmatrix}\,,\qquad  p_4^\mu=\frac{\sqrt{s}}{2}\begin{pmatrix} -1&-\cos\rho&-\sin\rho&\vec0_8\end{pmatrix}\,,\\
      q^\mu&=\frac{\sqrt{s}}{2}\begin{pmatrix} 1&\cos\theta&\sin\theta\cos\phi&\sin\theta\sin\phi\vec n_8\end{pmatrix}\,,
   }
   where $\vec n_8$ is the unit eight-vector and the scattering angle is
   \es{rho}{
   \cos\rho=\frac{t-u}{s}\,.
   }
   After changing to these variables, the measure in \eqref{cut} becomes
   \es{cutmeas}{
  dq^{11} \delta^{(+)}(q^2) \delta^{(+)}((p_1+p_2-q)^2)=\frac{s^{\frac72}}{2^7}\sin^8\theta|\sin^7\phi\,|d\theta\, d\phi \,d^8\vec n_8\delta(\vec n_8^2-1)\,,
   }
   where the integral over the seven-dimensional sphere gives $\vol(S^7)=\pi^4/3$.  The Mandelstam variables for the two internal tree level terms become
   \es{mands}{
  & t'=2p_1\cdot q=-\frac{s}{2}(1-\cos\theta)\,,\qquad  t''=-2p_4\cdot q=-\frac s2(1+\cos\theta\cos\rho+\sin\theta\cos\phi\sin\rho)\,,\\
   & u'=2p_2\cdot q=-\frac{s}{2}(1+\cos\theta)\,,\qquad  u''=-2p_3\cdot q=-\frac s2(1-\cos\theta\cos\rho-\sin\theta\cos\phi\sin\rho)\,.\\
   }
   Putting all these ingredients together, and normalizing by the tree level supergravity amplitude \eqref{anyD}, we get the expression \eqref{cut2} in the main text. As a check on the normalization of this formula, we can apply it to $\cA_{R|R}$ to get
   \es{discRR}{
   \frac{\text{Disc}_s\cA_{R|R}}{\cA_{R}}=&\frac{\pi  s^{5/2}}{11520 t^{7/2} (s+t)^{7/2}}\Big[2 i s \sqrt{t} \sqrt{s+t} \left(173 s^2 t^2+80 s^3 t+15 s^4+186 s t^3+93 t^4\right)-15 \pi  t^6\\
   &+15 i s \left(15 s^3 t^2+20 s^2 t^3+6 s^4 t+s^5+15 s t^4+6 t^5\right)
   \log \frac{-2 \sqrt{t} \sqrt{s+t}+s+2 t}{s}\Big]\,,
   }
   which exactly matches the $s$-channel discontinuity of \eqref{ARR}.

   \section{Details on CFT data extraction}

In this appendix we give details about the methods we used to extract CFT data. We first describe how the inversion integrals for the DD's presented in the main text can be derived from large spin perturbation theory. Then we describe the form of the small $z$ DD's for each 1-loop amplitude in each theory. Lastly, we describe the projection method that we used to extract CFT data from the Mellin amplitudes.

\subsection{Derivation of inversion integral}
\label{largeSpinApp}

The problem we want to solve is the following. Given the singular piece of the following sum
\begin{equation}
\sum_{\ell} \lambda_{\cB[02]_\ell}^2 C_{\cB[02]_\ell} \bar z^4g_{\ell+11,\ell+1}^{0,-2}(\bar z) =sing(\bar z) + \text{regular}\,,
\end{equation}
what are the $ \lambda_{\cB[02]_\ell}^2$ that reproduce such a singular part. In order to answer this question, we need to recall two facts. First, the collinear conformal blocks are eigenfunctions of a Casimir operator 
\begin{equation}
{\cal D} \bar z^4g_{\ell+11,\ell+1}^{0,-2}(\bar z) = J^2 \bar z^4g_{\ell+11,\ell+1}^{0,-2}(\bar z),~~~~{\cal D} =(1-\bar z)\bar z(2 \bar \partial+ \bar z \bar \partial^2)\,,
\end{equation}
with eigenvalue $J^2=(\ell+5)(\ell+6)$. Second, we can explicitly perform the above sum for GFFT with $(\lambda^{(0)}_{\cB[02]_\ell})^2$ given in \eqref{treedata2} and split the answer into a singular part and a regular part:
\begin{equation}
\sum_{\ell} (\lambda_{\cB[02]_\ell}^{(0)})^2C_{\cB[02]_\ell} \bar z^4g_{\ell+11,\ell+1}^{0,-2}(\bar z) = sing^{(0)}(\bar z) + \text{regular} \,.
\end{equation}
The singular part includes terms which are divergent as $\bar z \to 1$, but also terms that become divergent upon the application of the Casimir operator ${\cal D}$ a finite number of times. An example of such term is for instance $(1-\bar z)\log^2(1-\bar z)$. Since each conformal block is regular, and remains so after acting on it with the Casimir operator, these singular terms can only arise from an infinite sum over conformal blocks. For the case at hand we obtain
\begin{equation}
sing^{(0)}(\bar z) =  \frac{1}{6}\frac{1}{(1-\bar z)^3} - \frac{7}{9} \frac{1}{(1-\bar z)^2} + \frac{29}{18}\frac{1}{1-\bar z} \,.
\end{equation}
In order to show how the idea of large spin perturbation theory works, let's consider a simple example where $sing(\bar z) = \frac{1-\bar z}{\bar z^2} \log^2(1-\bar z)$. One can explicitly check that
\begin{equation}
\left( -\frac{1}{144} {\cal D}^4 + \frac{5}{36} {\cal D}^3-\frac{3}{4} {\cal D}^2+ {\cal D} \right)sing(\bar z) = sing^{(0)}(\bar z)\,,
\end{equation}
up to regular terms which are not important, so we would find that\\ $\lambda_{\cB[02]_\ell}^2 =(\lambda^{(0)}_{\cB[02]_\ell})^2 \left(  -\frac{1}{144} J^8 + \frac{5}{36} J^6-\frac{3}{4} J^4+ J^2 \right)^{-1} $, if we assume that the answer is analytic in spin. It turns out the same procedure works systematically around $\bar z=1$ for any generic singular part, which allows to find  $\lambda_{\cB[02]_\ell}^2 $ as a perturbative series around large spin, to all orders. Let us now assume we have a generic singularity of the form
\begin{equation}
\label{eqnh}
\sum_{\ell} \lambda_{\cB[02]_\ell}^2 C_{\cB[02]_\ell} \bar z^4g_{\ell+11,\ell+1}^{0,-2}(\bar z)  = h(\bar z) \log^2(1-\bar z)\,,
\end{equation}
and that there exist a Kernel $K(\ell,\bar z)$ such that
\begin{equation}
\lambda_{\cB[02]_\ell}^2  = 4\pi^2 \int_0^1 d\bar z K(\ell,\bar z)h(\bar z)\,,
\end{equation}
where for simplicity we assume $h(\bar z)$ has a double zero at $\bar z=1$ (the case with a single zero was just analysed above). Then acting on both sides of (\ref{eqnh}) with the Casimir we conclude
\begin{equation}
\label{eqnh2}
\sum_{\ell} J^2 \lambda_{\cB[02]_\ell}^2 C_{\cB[02]_\ell} \bar z^4g_{\ell+11,\ell+1}^{0,-2}(\bar z) = {\cal D} \left( h(\bar z) \log^2(1-\bar z) \right) = {\cal D}(h(\bar z)) \log^2(1-\bar z)\,,
\end{equation}
where the second equality works up to regular terms. But this implies
\begin{equation}
\int_0^1 d\bar z J^2 K(\ell,\bar z)h(\bar z) = \int_0^1 d\bar z K(\ell,\bar z) {\cal D} h(\bar z) = \int_0^1 d\bar z h(\bar z) {\cal D}^\dagger K(\ell,\bar z) \,,
\end{equation}
so that the Kernel should satisfy the following constraints:
\begin{itemize}
\item It is an eigenfunction of the adjoint Casimir operator ${\cal D}^\dagger$ with eigenvalue $J^2$:
\begin{equation}
{\cal D}^\dagger  K(\ell,\bar z) = J^2 K(\ell,\bar z),~~~~~{\cal D}^\dagger = (1-\bar z)\bar z^2 \bar \partial^2-2\bar z(2\bar z-1) \bar \partial -2\bar z
\end{equation}
with the correct boundary conditions at $\bar z=0$, such as to make the inversion integral convergent for large enough spin. 
\item Its normalised such that we recover the expected GFFT value $(\lambda^{(0)}_{\cB[02]_\ell})^2$ given in \eqref{treedata2}. Equivalently we could have used the example analysed above. 
\end{itemize}
These conditions fix the Kernel used in the body of the paper for $\lambda^2_{\cB[02]_\ell}$, and a very similar argument gives the Kernel for the anomalous dimension. 

\subsection{Double-discontinuities at small $z$}
\label{smallDD}

For $A_{N-1}$, the relevant DD for computing $( \lambda^{R|R}_{{\cal B}[02]_\ell})^2$ follows from the resummation of the leading terms in \eqref{sing} and we get
\begin{eqnarray}
h_\text{no-log}^{A,R|R}(\bar z) = \frac{ Q^{(8)}_{A,R|R}(\bar z)}{ (1-\bar z)^{5/2} \bar z^{13/2}}\arcsin\left(\sqrt{1-\bar z}\right) +  \frac{ Q^{(16)}_{A,R|R}(\bar z)}{(1-\bar z)^2 \bar z^{6}}\,,
\end{eqnarray}
where the polynomials have degree 8 and 16, respectively, and are given in the attached \texttt{Mathematica} notebook, as will be all the similar polynomials below. The integrand in the inversion integral \eqref{inversion} then goes like $\bar z^{\ell-\frac32}$ for small $\bar z$, so it converges for $\ell\geq1$ as claimed. For $( \lambda^{R|R^4}_{{\cal B}[02]_\ell})^2$, the DD has the structure
\begin{eqnarray}
h^{R|R^4,A}_\text{no-log}(\bar z) = \frac{Q^{(12)}_{R|R^4,A}(\bar z)}{(1-\bar z)^{7/2} \bar z^{19/2}}\arcsin\left(\sqrt{1-\bar z}\right) +  \frac{Q^{(23)}_{2,R|R^4,A}(\bar z)}{(1-\bar z)^3 \bar z^{9}}\,.
\end{eqnarray}
The integrand in the inversion integral \eqref{inversion} then goes like $\bar z^{\ell-\frac92}$ for small $\bar z$, so it converges for $\ell\geq5$ as claimed. For anomalous dimensions $\gamma_{8,\ell}^{R|R}$ of the lowest twist long multiplet, the relevant DD now comes from the term with $z\log z$ in \eqref{sing}, which gives
\begin{eqnarray}
h^{A,R|R}_{\log }(\bar z) = \frac{Q^{(9)}_{A,R|R}(\bar z)}{(1-\bar z)^{5/2} \bar z^{15/2}}\arcsin\left(\sqrt{1-\bar z}\right) +  \frac{Q^{(4)}_{A,R|R}(\bar z)}{\bar z^{5}} \log \bar z +  \frac{Q^{(18)}_{A,R|R}(\bar z)}{(1-\bar z)^2 \bar z^{7}} \,.
\end{eqnarray}
 The integrand in the inversion integral \eqref{hat} then goes like $\bar z^{\ell-\frac52}$ for small $\bar z$, so it converges for $\ell\geq2$ as claimed. For $\gamma_{8,\ell}^{R|R^4}$,  the relevant DD is now 
\begin{eqnarray}
h^{R|R^4}_{{\log }}(\bar z) =\frac{Q^{(13)}_{R|R,A}(\bar z)}{(1-\bar z)^{7/2} \bar z^{21/2}}\arcsin\left(\sqrt{1-\bar z}\right) +  \frac{Q^{(25)}_{R|R,A}(\bar z)}{(1-\bar z)^3 \bar z^{10}} \,.
\end{eqnarray}
The integrand in the inversion integral \eqref{hatR4} then goes like $\bar z^{\ell-\frac{11}{2}}$ for small $\bar z$, so it converges for $\ell\geq6$ as claimed. 

The extraction of CFT-data for $D_{N}$ theories works in exactly the same way, except now we use the resummation of the $D_N$ slices, and the resulting expressions are more complicated. For $( \lambda^{R|R}_{{\cal B}[02]_\ell})^2$, the DD is
\es{DDD}{
h^{D,R|R}_\text{no-log}(\bar z) = &\frac{Q^{(15)}_{D,R|R}(\bar z)}{(1-\bar z)^{5/2} (1+\bar z)^{33/2}}\text{arctanh}\left(\sqrt{1-\bar z^2}\right) \\
&+  \frac{Q^{(32)}_{D,R|R}(\bar z)}{(1-\bar z)^2 \bar z^6 (1+\bar z)^{16}} +\frac{Q^{(8)}_{D,R|R}(\bar z)}{(1-\bar z)^{5/2} \bar z^{13/2}}\arcsin\left(\sqrt{1-\bar z}\right) \,,
}
and the behaviour around $\bar z=0$ is exactly the same as for the $A_{N-1}$ theories, hence the integral converges for $\ell\geq1$. For $( \lambda^{R|R^4}_{{\cal B}[02]_\ell})^2$, the DD is
\es{DDD2}{
h^{R|R^4,D}_\text{no-log}(\bar z) =&\frac{Q^{(21)}_{R|R^4,D}(\bar z)}{(1-\bar z)^{7/2} (1+\bar z)^{45/2}}\text{arctanh}\left(\sqrt{1-\bar z^2}\right) \\
&+  \frac{Q^{(45)}_{R|R^4,D}(\bar z)}{(1-\bar z)^3 \bar z^9 (1+\bar z)^{22}} + \frac{Q^{(12)}_{R|R^4,D}(\bar z)}{(1-\bar z)^{7/2} \bar z^{19/2}}\arcsin\left(\sqrt{1-\bar z}\right) \,,
}
and the integrals converge for $\ell\geq5$ as before. For anomalous dimensions $\gamma_{8,\ell}^{R|R}$ for $D_N$, we get the DD
\es{anomDDD}{
h^{D,R|R}_{\log}(\bar z) &=\frac{Q^{(9)}_{D,R|R}(\bar z)}{(1-\bar z)^{5/2} \bar z^{15/2}}\arcsin\left(\sqrt{1-\bar z}\right) +  \frac{Q^{(4)}_{D,R|R}(\bar z)}{\bar z^{5}} \log \bar z +  \\
&  + \frac{Q^{(25)}_{D,R|R}(\bar z)}{(1-\bar z)^{5/2} (1+\bar z)^{37/2}\bar z^5}\text{arctanh}\left(\sqrt{1-\bar z^2}\right)  +  \frac{Q^{(36)}_{D,R|R}(\bar z)}{(1-\bar z)^{5/2} \bar z^{5}(1+\bar z)^{37/2}} \,,
}
and the inversion integral again converges for $\ell\geq2$. Finally, for $\gamma_{8,\ell}^{R|R^4}$ we have the relevant DD:
\es{anomDDD2}{
h^{R|R^4,D}_{\log}(\bar z) &=   \frac{Q^{(22)}_{R|R^4,D}(\bar z)}{(1-\bar z)^{7/2} (1+\bar z)^{49/2}}\text{arctanh}\left(\sqrt{1-\bar z^2}\right)  \\
&  +  \frac{Q^{(49)}_{R|R^4,D}(\bar z)}{(1-\bar z)^{3} \bar z^{10}(1+\bar z)^{24}}+\frac{Q^{(13)}_{R|R^4,D}(\bar z)}{(1-\bar z)^{7/2} \bar z^{21/2}}\arcsin\left(\sqrt{1-\bar z}\right)\,,
}
and the inversion integral converges from $\ell\geq6$ as before. 

\subsection{The projection method}
\label{extraction}

We now describe how to extract the CFT data from the Mellin amplitude in Section \ref{reducedMellinAmplitude} using the projection method, as was used in the similar context of 3d ABJM theory in \cite{Chester:2018lbz}.

Since we will be mainly interested in extracting the CFT data of leading twist operators, it is convenient to expand out the 6d conformal blocks in the lightcone block expansion and use orthogonality relations for hypergeometric functions to project onto a specific spin. To give a concrete example we look at twist 6 non-chiral algebra operators $\cD[04]$ and $\cB[02]_\ell$ (see Table \ref{moreMult}) and extract their OPE coefficients at order $c^{-2}$. The ingredients that we need are as follows:
\begin{itemize}
\item The collinear block expansion of the reduced one-loop correlator at order $U^4$. This is essentially obtained from the last line in \eqref{RR} by expanding out the 6d blocks in lightcone blocks to zeroth order. We get
\begin{align}\label{ltcoll}
\cH^{R|R}\bigg |_{U^4}=(\lambda_{\cD[04]}^{R|R})^2 C_{\cD[04]} g^{0,-2}_{10,0}(V)+\sum_{\ell\in\text{Odd}}(\lambda_{\cB[02]_\ell}^{R|R})^2C_{\cB[02]_\ell }g^{0,-2}_{\ell+11,\ell+1}(V)\,,
\end{align}

\item To project onto a specific spin we note the following identity from \cite{Heemskerk:2009pn,Heemskerk:2010ty}. Let $f(a,b,c;z)=z^a {}_2F_1(a,b,c;z)$ and $\tilde{f}(a,b,c;z)=z^{1+a-c} {}_2F_1(1+a-c,1+b-c,2-c;z)$. Then we have the following orthogonality relation
\begin{align}
\frac{1}{2\pi i}\int dz &~f(a_0+n,b_0+n,c_0+n;z)\no\\
&\cdot \tilde{f}(a_0+m,b_0+m,c_0+2m;z)(z-1)^{a_0+b_0-c_0}z^{c_0-2a_0-2}=\delta_{m,n}~.
\end{align}
Adapting to the case at hand we find that to project onto an operator of dimension $\Delta'$ and spin $j'$ all we need to do is to simply multiply by the following quantity
\begin{align}
\label{projectorG}
\cP_{\Delta',\ell'}(V)=&-\frac{1}{V(1-V)^{\ell'+1}} \\
& \, _2F_1\left(-\frac12(\Delta'+\ell'+4),-\frac12(\Delta'+\ell'+4)+1;-(\Delta'+\ell'+4)+2;1-V\right)\no
\end{align}
and take the residue at $V=1$.
\item Next we calculate the $U^4$ term of the reduced one-loop correlator in \eqref{RR}. This is obtained from $M^{R|R}$ \eqref{MellinRR} by selecting the $s=6$ pole which contributes to the $s$-integral in \eqref{mellinH}. This gives us
\be
-U^4\sum_{m=4}^\infty c^A_m\int_{-i\infty}^{i \infty}\frac{dt}{2\pi i}\frac{\pi ^2 (m-1) (t-6)^2 (t-4)^2 (t-2)^2 t^2 (t+2)^2 V^{\frac{t}{2}-3}}{1024 (2 m-t) (2 m+t-4)\sin^2(\pi t/2)}
\ee
where the coefficients $c^A_m$ are given in \eqref{cmhatlow} and \eqref{cmhat}. Then we perform the $t$-integral. Although this can be done explicitly for generic value of $m$ the result is a rather complicated function $fun(V)$ of $V$ that we refrain from presenting here. This function contains the CFT data of $\cD[04]$ and $\cB[02]_\ell$ for all spin $\ell$.
\item The real simplification comes when we apply the projector \eqref{projectorG} on $fun(V)$ and as a result we find a numerical sum over $m$. As an example for $\cD[04]$ we have $\Delta'=6$ and $\ell'=0$. With these values when we project on $fun(V)$ we get the following convergent sum
\begin{align}\label{spin0sum}
&\sum_{m=4}^\infty \bigg[\frac{1}{105} \big(-210 m^9+1365 m^8-2555 m^7-595 m^6+5712 m^5-2975 m^4-3085 m^3\no\\
&+1905 m^2-462 m-780\big) +2 (m-3)^2 (m-2)^2 (m-1)^2 m^2 (m+1)^2 \psi ^{(1)}(m-3)\bigg] c^A_m~.
\end{align}
The result of this sum is to be equated with $(\lambda_{\cD[04]}^{R|R})^2 C_{\cD[04]} $ which comes from applying the projector $\cP_{6,0}(V)$ on the right side of \eqref{ltcoll}. After performing the sum we get the result for the term in \eqref{spin0A} that does not depend $B_4^{R|R}$.

These steps can be repeated for obtaining the OPE of $\cB[02]_\ell$. For instance for $\ell=1$ we find the following result in the $A_{N-1}$ theory
\begin{align}
&( \lambda^{R|R}_{{\cal B}[02]_1})^2= \sum_{m=4}^\infty \bigg[-\frac{1}{693}(m+1) \big(2310 m^{10}-21945 m^9+83650 m^8-156520 m^7\no\\
&+124033 m^6+44471 m^5-173090 m^4+134840 m^3-35403 m^2+1434 m+1260\big)  \no\\
&+\frac{10}{33}(m-3)^2(m-2)^2(m-1)^2 m^2 (m+1)^2 \left(11 m^2-22 m+17\right)\psi ^{(1)}(m-3)\bigg]c_m^A~,\no\\
&=\frac{1314669460}{231}-\frac{6738525487050 \pi ^2}{11685817}~,
\end{align}
which matches perfectly with the result from the inversion integral in Section \ref{CFTData}. Here we remark that it is important to do the $t$-integral before doing the projection since the reverse procedure leads to a divergent sum in $m$.
\item To compute the anomalous dimensions of leading twist unprotected operators, the steps are very similar. There is however one new ingredient. We now look at the $\log U$ pieces in block expansion of the one-loop correlator \eqref{RR}. This involves a piece proportional to the derivative of the conformal blocks $\partial_\Delta G_{\Delta,\ell}$. When we expand it out in collinear blocks to just the leading power of $U$, hit it with the projector $\cP_{\Delta',\ell'}$ and take the residue at $V=1$, we pick all $G_{\Delta,\ell}$ such that $\ell<\ell'-1$. We need to subtract these contributions from the sum in $m$ and $n$ (which appears in the expression of the one-loop Mellin amplitude) to obtain the final result for the one-loop anomalous dimension.
\end{itemize}
In this way we have obtained the leading twist spin 0, 2,4 and 6 anomalous dimension at order $c^{-2}$ for both $A_{N-1}$ and $D_N$ theories. For spin $>0$, we find perfect agreement with the result obtained from the inversion integral quoted in Section \ref{CFTData}. There is a freedom of adding contact terms to the $R|R$ Mellin amplitude which will affect the spin 0 CFT data. A rather long calculation yields the result for the leading twist, spin 0 anomalous dimension for both the $A_{N-1}$ and $D_N$ theories as shown in \eqref{spin0A} in the main text.

\bibliographystyle{JHEP}
\bibliography{6ddraft}

\end{document}